\documentclass[11pt]{article}
\usepackage[margin=1in]{geometry}
\usepackage{amsmath,amssymb}
\usepackage{graphicx}
\usepackage{hyperref}
\usepackage{booktabs}
\usepackage{enumitem}
\usepackage{xcolor}
\usepackage{appendix}
\newcommand{\bl}[1]{{\color{blue}#1}}

\title{\vspace{-1em}Cycles Without End: An Ekpyrotic Braneworld Bounce with a Kinetically Coupled
Entropic Mechanism\vspace{-0.5em}}
\date{}

\author{
Rikpratik Sengupta$^{1}$\\[4pt]
\small $^{1}$Department of Physics, Indian Institute of Technology Kanpur, Kalyanpur 208016, India.
}

\begin{document}
\maketitle
\vspace{-2.5em}

\begin{abstract}
\noindent Cyclic cosmologies replace inflation's single creation event with an eternal sequence
of smooth turnarounds, but three problems have historically stood in their way: singular bounces,
the blue tilt generic to single-field ekpyrotic contraction, and Tolman's observation that entropy
production should make successive cycles longer and larger without bound. We address all three
within one explicit model. A Randall--Sundrum brane with a timelike extra dimension
supplies a bounce that requires no violation of the null energy condition by the matter on the
brane, and a single scalar field's potential carries the universe from a thawing dark-energy
plateau through a sign-changing turnaround into a BKL-safe ekpyrotic contraction and back, with no
curvature or negative cosmological constant required anywhere. The density is confined to
$0\le\rho\le\rho_c$ throughout, the recollapse is driven by the negative potential, and we prove
that it is unavoidable for the potential used. We show that no correction internal to the brane's
Friedmann equation, and no straight two-field trajectory, can rescue the spectral index
of the adiabatic mode from the generic ekpyrotic result $n_s\approx3$, using the full
closed system of brane perturbation equations including the projected Weyl fluid, whose effect on
modes that leave the horizon at $\rho\ll\rho_c$ is shown to be suppressed by $\rho/\rho_c$. We
then build the minimal completion that can rescue the tilt: a kinetically coupled entropy
field, following the general mechanism of Ijjas, Lehners, and Steinhardt, whose coupling steepness
has a closed-form, exact-scale-invariance value for this model's ekpyrotic potential, with only a
percent-level offset needed to match the measured tilt $n_s=0.9649$. This fixes, in closed form and
with no further freedom, the local non-Gaussianity ($f_{\rm NL}\approx-0.02$) and the
tensor-to-scalar ratio ($r\approx7.6\times10^{-26}$). Matching the observed amplitude fixes the
conversion coupling and the brane tension to unremarkable values ($\lambda\approx5.7\times10^{-3}$,
$\rho_c^{1/4}=10^{13}\,$GeV). Finally, because the brane construction is spatially flat, Tolman's
entropy argument does not bound the number of cycles here. We derive the net number of e-folds
per cycle, $N_{\rm cyc}\gtrsim77$-$80$, needed to dilute the entropy produced (dominated by
supermassive black holes), show that ekpyrotic contraction followed by expansion already
supplies $16$-$35$ of them, and conclude that the dark-energy phase must supply
$N_{\rm DE}\gtrsim42$-$64$ e-folds depending on the post-bounce history. The model does not
contain a mechanism that produces the hot radiation era after the bounce, and its predictions
concern the spectrum generated during the ekpyrotic phase.
\end{abstract}

\tableofcontents

\section{Introduction}
\label{sec:intro}

The standard cosmological paradigm places a single inflationary epoch before the hot big bang,
solving the horizon and flatness problems and seeding structure from quantum fluctuations
stretched to cosmological scales. Cyclic and ekpyrotic cosmologies pursue an alternative in which
these problems are solved instead by a slow, gentle \emph{contraction}, repeated without end, with
no beginning and, in principle, no final state. The ekpyrotic proposal
\cite{KhouryOvrutSteinhardtTurok2001} showed that a phase of contraction driven by a steep,
negative scalar-field potential is a powerful smoothing and flattening mechanism. Unlike ordinary
contraction, which is a chaotic, Belinski--Khalatnikov--Lifshitz (BKL) unstable approach to a
crushing singularity dominated by anisotropy, contraction sourced by a fluid or field with
equation-of-state parameter $w\gg1$ makes the anisotropy energy density, which redshifts as
$a^{-6}$, utterly negligible relative to the field's own energy density, which redshifts far more
slowly. Steinhardt and Turok combined this with a big-crunch/big-bang transition to build the
cyclic universe \cite{SteinhardtTurok2002Science,SteinhardtTurok2002PRD}, in which the entire
cosmic history repeats indefinitely: radiation and matter domination, the current
dark-energy-dominated acceleration, a future turnaround, ekpyrotic contraction, and a bounce back
into expansion. Subsequent work sharpened nearly every part of this proposal. Steinhardt and Turok
later showed that a sufficiently long phase of the present cosmic acceleration is itself required
to smooth and flatten each cycle to the precision observed \cite{SteinhardtTurok2005}. Bounces that are free of ghost and gradient instabilities around the turning point were first
constructed in Galileon-type theories \cite{QiuEtAl2011,EassonSawickiVikman2011} and later in
cubic Horndeski models \cite{IjjasSteinhardt2016Nonsingular,IjjasSteinhardt2017Fully}. The
solution of Ref.~\cite{IjjasSteinhardt2016Nonsingular} was subsequently shown to begin with a
divergent speed of sound and to end with a vanishing one, so that it connects two strongly coupled
configurations within roughly $10^{2}$ Planck times of the bounce \cite{Dobre2018}, and the construction of
Ref.~\cite{IjjasSteinhardt2017Fully} was designed to remove such behavior. The use of these
bounces in an iterated, entropy-conscious cyclic cosmology is developed in
\cite{IjjasSteinhardt2019NoBB}.
It was shown that scale-invariant curvature perturbations require
either two fields, the entropic mechanism
\cite{Notari2002,Finelli2002,LehnersEtAl2007,LehnersSteinhardt2008,BuchbinderKhouryOvrut2007}, or a
single field with a specifically designed (``adiabatic'') running of the potential's steepness
\cite{KhouryySteinhardt2010,LiSteinhardt2010}. And a broader case has been made that inflation's
own successes can be reproduced, and its flaws (multiverse overproduction, unpredictivity,
trans-Planckian sensitivity) avoided, by non-inflationary smoothing mechanisms of exactly this
ekpyrotic type \cite{IjjasSteinhardt2018Inflation,IjjasLoebSteinhardt2013}. Reviews of the broader
non-Gaussianity, entropic-mechanism, and bounce-construction literature are given in
\cite{LehnersReview2010,BattefeldGeorges2015}, and the non-minimal ekpyrotic trispectrum is analyzed in \cite{FertigLehners2016}.

Three problems have shadowed every concrete realization of this program from the start. First,
the bounce itself: converting a contracting to an expanding universe in general relativity
requires violating the null energy condition (NEC) at least momentarily, forcing either exotic
matter or careful matched-asymptotic treatments across a genuine curvature singularity, with the
attendant risk that quantum-gravitational physics at the singularity is doing uncontrolled work.
Braneworld constructions offer a way out that requires no violation of the NEC by the matter on the brane (Section~\ref{sec:nec}). On a
Randall--Sundrum brane with a timelike extra dimension, the induced Friedmann equation is modified
at high density in a way that produces a bounce purely from the brane's own geometry
\cite{ShtanovSahni2003,SahniShtanov2002}, later adopted explicitly as the bounce mechanism in
hysteresis-type cyclic constructions \cite{SahniToporensky2012,SahniShtanovToporensky2015}. Second,
the spectral tilt: the simplest, single-field realization of ekpyrotic contraction generates a
badly blue spectrum, $n_s\to3$ as the field's steepness grows \cite{Lyth2002}, in stark
contradiction with the observed slight red tilt $n_s=0.9649\pm0.0042$ \cite{Planck2018Params}.
Third, and oldest, is Tolman's 1934 observation \cite{Tolman1934} that in a closed, oscillating
universe, entropy produced during each cycle, by dissipation, particle production, and structure
formation, must persist into the next. This forces each successive cycle to reach a larger maximum
radius and longer period than the last, precluding a truly eternal, self-similar sequence of
cycles.

This paper builds one explicit model, integrates it through its full sequence of phases with no
slow-roll or attractor approximation in the background dynamics, and addresses all three problems
within it. Section~\ref{sec:brane} sets up the Shtanov--Sahni brane bounce. Section~\ref{sec:matter}
covers the matter sector, a single scalar field whose potential is responsible for both the
present cosmic acceleration and its own eventual reversal into ekpyrotic contraction, with no
separate curvature or $\Lambda<0$ ingredient anywhere. Section~\ref{sec:bg} gives the resulting
background dynamics, verified through two complete cycles. Section~\ref{sec:tolman} addresses
Tolman's problem directly. It rests on a fact central to the modern cyclic-universe
literature~\cite{SteinhardtTurok2002PRD,SteinhardtTurok2005} but rarely made quantitative: a
\emph{spatially flat} cyclic universe has no finite total entropy to bound. We derive how many
e-folds of accelerated expansion each cycle's dark-energy phase must supply to keep the entropy
\emph{density} at the start of each ekpyrotic phase from secularly increasing.
Sections~\ref{sec:pert}--\ref{sec:brane_ns} take up the spectral tilt. We show that neither the simple $m=0$ brane nor its induced-gravity ($m\neq0$)
generalization can rescue the single-field result, for a clean, e-fold-counting structural reason.
Section~\ref{sec:entropic} builds the two-field entropic completion that does work: a kinetically
coupled entropy field built directly on the model's existing ekpyrotic branch, exactly solved, matched to the
observed tilt across the full observable window, converted to the observed curvature perturbation
by a minimal deformation of the model's own reflecting wall, and worked through to explicit,
closed-form predictions for the amplitude, non-Gaussianity, and tensor-to-scalar ratio.
Section~\ref{sec:discussion} discusses what this construction establishes, what it does not, and
concludes.

The scope of the construction should be stated at the outset. The primordial spectrum computed in
Sections~\ref{sec:pert}--\ref{sec:tensor} is generated during the ekpyrotic contraction, before the
bounce, and its tilt, amplitude, and non-Gaussianity are fixed by physics at densities far below
the brane tension. These quantities are carried into the expanding phase under two assumptions that
are stated where they are used: the super-horizon conservation of the adiabatic curvature
perturbation across the bounce, and the production of the post-bounce radiation from the energy of
the scalar field without the generation of additional isocurvature perturbations. The model does
not contain a mechanism that produces the hot radiation era after the bounce, and
Section~\ref{sec:reheat} states what such a mechanism would have to achieve. The reason for
computing the spectrum in a model of this kind is that the blue tilt of single-field ekpyrosis is
the standard obstruction to a viable cyclic cosmology. A cyclic model that cannot reproduce the
observed spectrum is not viable however the post-bounce universe is reheated, so the tilt, the
amplitude, and the tensor-to-scalar ratio are necessary tests of the construction. They are not
sufficient ones, and the model is not presented as a complete cosmology.

The analysis is organized so that its dependence on the brane construction is explicit. The
background results of Sections~\ref{sec:brane}--\ref{sec:bg} depend on the effective Friedmann
equation, Eq.~\eqref{eq:friedmann}, which is shared with loop quantum cosmology
(Section~\ref{sec:lqc}); the viability of the timelike extra dimension is discussed in
Section~\ref{sec:timelike}; the applicability of four-dimensional perturbation theory to
this brane is analyzed in Section~\ref{sec:brane_pert}; and the causal structure of an infinite
cyclic universe is treated in Section~\ref{sec:horizon}.

\section{Gravitational sector: braneworld bounce}
\label{sec:brane}

We use the Shtanov--Sahni braneworld~\cite{ShtanovSahni2003}: a codimension-one brane bounding a
five-dimensional bulk with timelike extra dimension ($\epsilon_{\rm brane}=-1$), zero
brane-localized curvature term ($m=0$, the Randall--Sundrum limit), spatially flat brane
($\kappa_{\rm FRW}=0$), and zero dark-radiation term ($C=0$). Spatial flatness is not a
simplifying choice made for convenience: it is what allows Section~\ref{sec:tolman}'s resolution
of Tolman's problem to go through at all. With the standard tuning $\Lambda_{\rm eff}=0$ between
the bulk cosmological constant and the brane tension, the induced Friedmann equation on the brane
is
\begin{equation}
H^2 = \frac{8\pi G_N}{3}\rho\Big(1-\frac{\rho}{\rho_c}\Big), \qquad
\rho_c \equiv \frac{8\pi G_N}{3}M^6 = 2|\sigma|,
\label{eq:friedmann}
\end{equation}
where $M$ is the five-dimensional Planck mass and $\sigma<0$ is the brane tension. We adopt units
$\kappa\equiv8\pi G_N/3=1$ for the numerical work of Sections~\ref{sec:matter}--\ref{sec:tolman},
so $H^2=\rho(1-\rho/\rho_c)$. The acceleration equation is
\begin{equation}
\dot H = -\frac32(\rho+p)\Big(1-\frac{2\rho}{\rho_c}\Big).
\label{eq:accel}
\end{equation}
Two structural properties make Eqs.~\eqref{eq:friedmann}--\eqref{eq:accel} the right foundation for
a numerically robust cyclic model. First, the Friedmann constraint is conserved identically given
only ordinary stress-energy conservation $\dot\rho=-3H(\rho+p)$:
\begin{equation}
\frac{d}{dt}\Big[H^2-\rho\Big(1-\frac\rho{\rho_c}\Big)\Big] = 2H\dot H - \dot\rho\Big(1-\frac{2\rho}{\rho_c}\Big) = 0
\label{eq:constraint}
\end{equation}
identically, substituting Eqs.~\eqref{eq:accel} and $\dot\rho=-3H(\rho+p)$. So if
Eq.~\eqref{eq:friedmann} holds at $t=0$ it holds at all $t$ along the exact solution. We integrate
$H$ directly via Eq.~\eqref{eq:accel} rather than $H=\pm\sqrt{\rho(1-\rho/\rho_c)}$ with a
hand-flipped branch at each turning point; this removes any sign ambiguity at $H=0$ and lets
bounces and turnarounds emerge from the dynamics rather than being imposed. Second, at
$\rho=\rho_c$, $H=0$ identically regardless of the equation of state, and
$\ddot a/a=\frac32(\rho+p)>0$ whenever $\rho+p>0$: true for any ordinary or ekpyrotic matter. This
is a generic, energy-condition-respecting bounce, requiring no exotic matter and no
singular matching. The nucleosynthesis-safety bound of~\cite{ShtanovSahni2003} requires
$|\sigma|\gtrsim(1\,{\rm MeV})^4$; we return to this bound with an explicit physical value in
Section~\ref{sec:amplitude}.

\subsection{Meaning of the bounce and the effective null energy condition}
\label{sec:nec}

The statement that the bounce requires no violation of the null energy condition refers to the
matter confined to the brane, for which $\rho+p=\dot\phi^2\ge0$. It is useful to state what the
effective four-dimensional description does. Writing Eq.~\eqref{eq:friedmann} as
$H^2=\kappa\rho_{\rm eff}$ with $\rho_{\rm eff}=\rho(1-\rho/\rho_c)$, and defining the effective
pressure by conservation of the effective source,
$p_{\rm eff}=p-(\rho/\rho_c)(2p+\rho)$, one finds
\begin{equation}
\rho_{\rm eff}+p_{\rm eff}=(\rho+p)\Big(1-\frac{2\rho}{\rho_c}\Big),
\label{eq:effnec}
\end{equation}
which is negative for $\rho>\rho_c/2$. The effective source therefore violates the null energy
condition in the high-density region, as any bounce of a spatially flat universe must in four
dimensions, while the brane matter does not. The origin of the effective violation is the
quadratic correction that descends from the bulk-brane geometry through the junction conditions, so
that no exotic matter and no singular matching are involved. The same structure appears in loop
quantum cosmology, as discussed next.

\subsection{Correspondence with loop quantum cosmology}
\label{sec:lqc}

For a spatially flat universe the effective Friedmann equation of loop quantum cosmology has the
form $H^2=(8\pi G/3)\rho(1-\rho/\rho_{\rm crit})$, where $\rho_{\rm crit}$ is of order the Planck
density \cite{SinghVandersloot2006,AshtekarPawlowskiSingh2006}. This coincides with
Eq.~\eqref{eq:friedmann} with $\rho_c\to\rho_{\rm crit}$. The coincidence is exact at the level of
the background and has useful consequences. In both theories the density is confined to
$0\le\rho\le\rho_c$, the Hubble rate is bounded by $H_{\max}^2=\kappa\rho_c/4$ in units where
$\kappa=8\pi G/3$, attained at $\rho=\rho_c/2$, and a turnaround in the scale factor can occur at
two places. At $\rho=\rho_c$ the turnaround is a bounce for any matter with $\rho+p>0$. At $\rho=0$ it
is a classical recollapse that requires a negative potential, because $\rho=0$ with
$\rho+p=\dot\phi^2>0$ implies $V=-\dot\phi^2/2<0$, a situation analyzed for LQC in
Ref.~\cite{SinghVandersloot2006}, where it is shown that cyclic evolution without
singularities is obtained when the potential is negative in one region and slightly positive in
another. The potential of Section~\ref{sec:matter} has this structure, with its positive thawing
plateau playing the role of the slightly positive branch.
The consequences of Sections~\ref{sec:matter}--\ref{sec:tolman} for the background, including the
turnaround and the BKL-safety analysis, therefore apply to an LQC-type bounce with $\rho_c$
reinterpreted. The correspondence does not extend to perturbations, whose equations differ in the
two frameworks, and the perturbative analysis of this paper is carried out for the brane
(Section~\ref{sec:brane_pert}).

\subsection{Status of the timelike extra dimension}
\label{sec:timelike}

The bulk of the Shtanov--Sahni brane has signature with two timelike directions, and it is
known that such bulks raise concerns about tachyonic and negative-norm modes, about causality, and
about the well-posedness of the bulk initial-value problem
\cite{Yndurain1991,DvaliGabadadzeSenjanovic1999}. Arguments that these pathologies are absent for
a non-compact timelike extra dimension with the matter localized on a brane have been given in
Ref.~\cite{IglesiasKakushadze2001}, topological criteria for the absence of ghosts in related
settings are discussed in Ref.~\cite{ArefevaVolovich1985}, and a timelike extra dimension has been
employed in other model-building contexts \cite{ChaichianKobakhidze2000}. These arguments do not
amount to a proof that the brane used here is free of instabilities, and no such proof is claimed.

The results of this paper depend on the construction in a limited and identifiable way. The
perturbative results of Sections~\ref{sec:pert}--\ref{sec:tensor} concern densities
$\rho\ll\rho_c$, where Eq.~\eqref{eq:friedmann} reduces to general relativity with corrections of
order $\rho/\rho_c$. The bounce enters only through Eq.~\eqref{eq:friedmann}, which is shared with
LQC. With the definition $\rho_c=\kappa M^6$ of Eq.~\eqref{eq:friedmann}, where $M$ is the
five-dimensional Planck mass, the fiducial brane tension $\rho_c^{1/4}=10^{13}\,$GeV adopted in
Section~\ref{sec:amplitude} corresponds, in reduced Planck units with $\kappa=1/(3M_{\rm pl}^2)$ for
$8\pi G=M_{\rm pl}^{-2}$, to
\begin{equation}
M=\big(3\rho_cM_{\rm pl}^2\big)^{1/6}\simeq7.5\times10^{14}\,{\rm GeV},\qquad
\frac{\rho_c^{1/4}}{M}\simeq1.3\times10^{-2},\qquad
\frac{H_{\max}}{M}\simeq1.6\times10^{-8},
\label{eq:Mfive}
\end{equation}
with $H_{\max}=\sqrt{\rho_c/12}/M_{\rm pl}\simeq1.2\times10^7\,$GeV. At the nucleosynthesis floor
$\rho_c^{1/4}=1\,$MeV the same relation gives $M\simeq1.6\times10^{4}\,$GeV. The whole evolution,
including the bounce, therefore takes place at energies and curvatures well below the
five-dimensional Planck scale, which is the regime in which a derivative expansion of the
induced brane dynamics is expected to apply. Any instability in the Kaluza--Klein sector would
have to act on this scale or be tied to the bulk structure itself, and its absence has to be
established within the five-dimensional theory. The conclusions of Sections~\ref{sec:pert}--%
\ref{sec:tensor} do not require the bounce to be realized by this particular brane. They
require that the ekpyrotic phase be described by a four-dimensional effective theory at
$\rho\ll\rho_c$ and that curvature perturbations be conserved across the bounce
(Section~\ref{sec:conversion}).

\section{Matter sector: a dark-energy-to-ekpyrosis crossover potential}
\label{sec:matter}

The matter sector is a single canonical scalar $\phi$, $\mathcal L=\frac12\dot\phi^2-V(\phi)$, so
that $\rho=\frac12\dot\phi^2+V(\phi)$, $p=\frac12\dot\phi^2-V(\phi)$,
$\ddot\phi+3H\dot\phi+V'(\phi)=0$. The potential is built from three smoothly-joined pieces (exact
form and parameters in Appendix~\ref{app:params}):
\begin{align}
V(\phi) &= V_{\rm plat}(\phi) + V_{\rm ekp}(\phi) + V_{\rm wall}(\phi), \label{eq:potential}\\
V_{\rm plat}(\phi) &= \frac{V_\Lambda}{\cosh^2\big((\phi-\phi_{\rm peak})/\phi_0\big)}, \\
V_{\rm ekp}(\phi) &= -V_0\,\sigma\!\Big(\frac{\phi-\phi_c}{\Delta_g}\Big)
\exp\big[c(\phi)(\phi-\phi_c)\big], \qquad
c(\phi) = c_1+(c_0-c_1)e^{-\max(\phi-\phi_c,0)/L}, \\
V_{\rm wall}(\phi) &= V_w\,\sigma\!\Big(\frac{\phi-\phi_w}{\Delta_w}\Big)\exp\big[c_w(\phi-\phi_w)\big],
\qquad \sigma(x)\equiv\tfrac12(1+\tanh x).
\end{align}
For $\phi\lesssim\phi_c$, $V\simeq V_{\rm plat}$ is a thawing plateau
($\epsilon_{\rm DE}\equiv\frac12(V'/V)^2\sim10^{-2}$, $w\to-1^+$): ordinary quintessence, not
inflation, sourcing a phase of accelerated expansion whose \emph{duration} (in real, un-compressed
units) is exactly the quantity Section~\ref{sec:tolman} needs to be long. Near $\phi_c$ the
potential descends and changes sign over a narrow gate $\Delta\phi\ll1$; as $\rho\to0^+$ while
$p=\frac12\dot\phi^2-V>0$ (since $V$ is by then large and negative), Eq.~\eqref{eq:accel} gives
$\dot H<0$ right at $H=0$: a genuine turnaround sourced purely by the field's equation of state,
with no curvature or negative-$\Lambda$ ingredient, realizing in a single field the mechanism
sketched for general fluids in \cite{SahniToporensky2012}. For $\phi>\phi_c$, $V_{\rm ekp}$ is a
steep negative exponential with designer running steepness $c(\phi)$ relaxing from $c_0$ (onset)
to $c_1<c_0$ (asymptotic), giving a slowly decreasing
$\epsilon(\phi)\equiv\frac12(V'/V)^2\simeq c(\phi)^2/2$. This is the same designer-running
philosophy pursued, for a different purpose (making $n_s$ itself run to unity), by
\cite{KhouryySteinhardt2010,LiSteinhardt2010}; we show in Section~\ref{sec:brane_ns} that a
BKL-safe running of this kind cannot by itself fix $n_s$, which is why
Section~\ref{sec:entropic} is needed. For $\phi>\phi_w$, a steep positive exponential $V_{\rm
wall}$ reflects the post-bounce kination phase back into deceleration and eventual re-thaw, playing
the role of the reflecting wall of the original cyclic construction
\cite{SteinhardtTurok2002PRD}. No potential minimum is used anywhere; there is no
oscillating-inflaton reheating step (Section~\ref{sec:reheat}).

\subsection{Shape of the potential}
\label{sec:potshape}

Figure~\ref{fig:potential} shows $V(\phi)$ for the parameters of Appendix~\ref{app:params}. For
$\phi\lesssim\phi_c$ the potential is the positive thawing plateau of height $V_\Lambda=10^{-3}$
centered at $\phi_{\rm peak}=-30$, on which the field is slowly rolling toward larger $\phi$. At
$\phi_c=10$ the switch function $\sigma$ turns on the ekpyrotic branch over the gate width
$\Delta_g=0.05$, and $V$ changes sign at $\phi\simeq9.75$, five gate widths below $\phi_c$, where the descending
exponential overtakes the plateau value. The
potential is then a steep negative exponential whose steepness relaxes from $c_0=4.2$ to
$c_1=3.2$ over $L=3$ units of field. The positive wall is positioned beyond the range of field
values reached before the bounce in the parameter set of Appendix~\ref{app:params}, which places
$\phi_w=18$ for the perturbative sector. In the numerical background of Section~\ref{sec:bg} the
reflecting wall sits closer to the gate, as described there. The bounce takes place at
$\rho=\rho_c$, which for $\rho_c=50$ is reached for $\phi-\phi_c$ of order unity.

\begin{figure}[h]
\centering
\includegraphics[width=\textwidth]{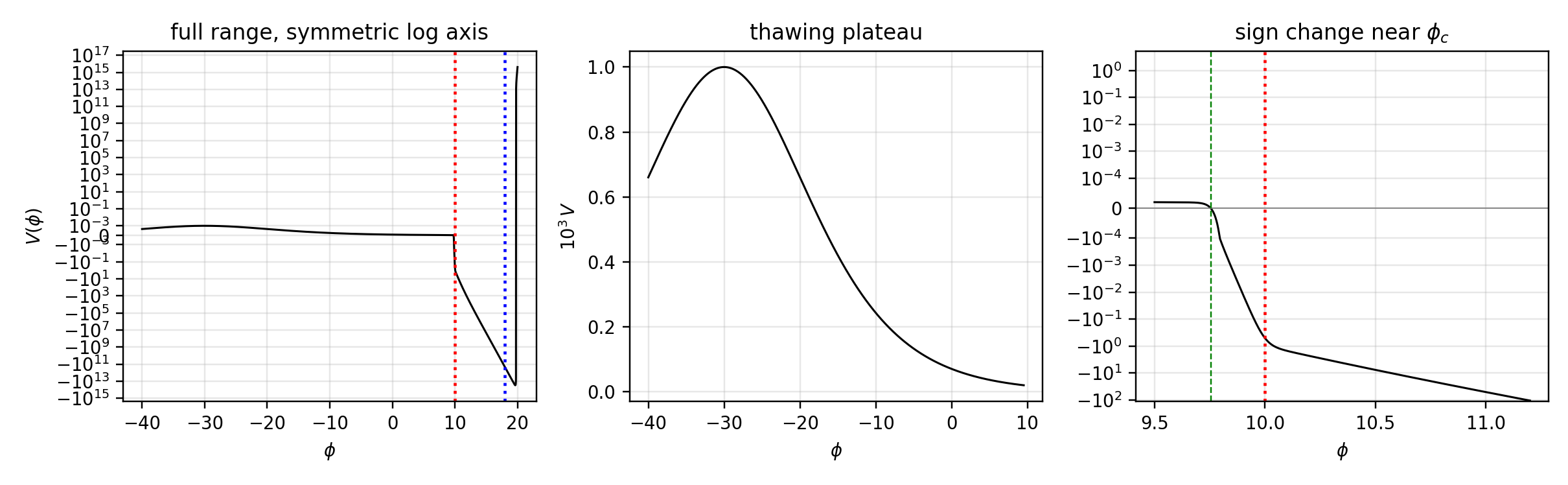}
\caption{The potential $V(\phi)$ of Eq.~\eqref{eq:potential} for the parameters of
Appendix~\ref{app:params}. Left: full range on a symmetric logarithmic axis, with $\phi_c$ (red,
dotted) and $\phi_w$ (blue, dotted). Center: the thawing plateau. Right: the sign change at the
gate near $\phi_c$, with the zero crossing marked by the dashed green line.}
\label{fig:potential}
\end{figure}

\subsection{Range of the density, the turnaround, and the inevitability of recollapse}
\label{sec:turnaround}

The recollapse in this model is driven by the negative potential. The following statements make
this precise and show that the energy density never becomes negative.

\emph{The density is confined to an interval.} Since $H$ is real, Eq.~\eqref{eq:friedmann}
requires $\rho(1-\rho/\rho_c)\ge0$, that is $0\le\rho\le\rho_c$ along any solution. Because
Eq.~\eqref{eq:friedmann} is conserved by the dynamics, Eq.~\eqref{eq:constraint}, this is a property of the exact
solution and not of a particular integration.

\begin{figure}[h]
\centering
\includegraphics[width=\textwidth]{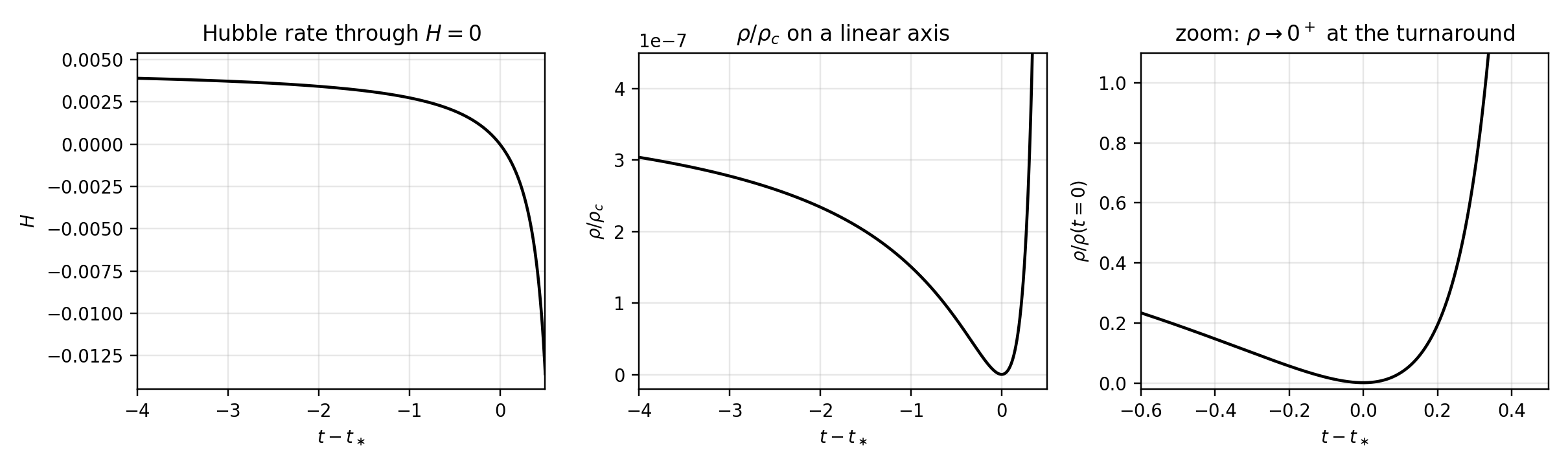}
\caption{The first turnaround of the exact background integration of Section~\ref{sec:bg}.
Left: $H(t)$ crossing zero with negative slope. Center: $\rho/\rho_c$ on a linear axis, which
decreases to a quadratic minimum and rises again. Right: zoom on $\rho/\rho(t=0)$ near the
turnaround. The residual minimum of order $10^{-3}$ of the initial density is the numerical
constraint residual of the direct integration of Eq.~\eqref{eq:accel}; the exact minimum is
zero. Time is measured from the crossing, $t_\ast\simeq3.2\times10^{3}$.}
\label{fig:turnaround}
\end{figure}

\emph{Where the expansion can stop.} At $H=0$ either $\rho=\rho_c$ or $\rho=0$. At $\rho=\rho_c$,
Eq.~\eqref{eq:accel} gives $\dot H=+\tfrac32(\rho+p)>0$ for matter with $\rho+p>0$, so the
contraction is reversed. At $\rho=0$, Eq.~\eqref{eq:accel} gives
$\dot H=-\tfrac32(\rho+p)=-\tfrac32\dot\phi^2<0$, so the expansion is reversed into contraction.
For a canonical scalar field $\rho=0$ means $\tfrac12\dot\phi^2=-V$, so a turnaround at $\rho=0$
with the field in motion requires $V<0$, and
\begin{equation}
\dot H\big|_{\rho=0}=3V(\phi_\ast)<0,\qquad
\frac{\ddot a}{a}\Big|_{\rho=0}=3V(\phi_\ast)<0
\label{eq:turnH}
\end{equation}
in units $\kappa=1$. For the parameters of Appendix~\ref{app:params} the field has $V(\phi_\ast)=-2.3\times10^{-3}$
at the first turnaround, which gives $\dot H=-6.9\times10^{-3}$, in agreement with the numerical
slope of $H(t)$ in Figure~\ref{fig:turnaround}.

\emph{The density touches zero and returns.} Near the turnaround $\rho\simeq H^2/\kappa$, and
$H$ passes through zero with the nonvanishing slope of Eq.~\eqref{eq:turnH}, so $\rho\propto
(t-t_\ast)^2$. The density has a quadratic minimum equal to zero at $t_\ast$ and rises again, as
dictated by $\dot\rho=-3H(\rho+p)$ with $H<0$ after the turnaround. It does not become negative.
That $\rho$ reaches exactly zero at $H=0$ is not a tuning but a consequence of the conserved
constraint, since $H^2=\rho(1-\rho/\rho_c)$ vanishes with $H$. The logarithmic axis of Fig.~\ref{fig:twocycles}
cannot display a zero, and Figure~\ref{fig:turnaround} shows the same evolution on a linear axis.

\emph{Recollapse is inevitable for this potential.} Suppose that on the branch traversed by the
field after the plateau, $V'(\phi)\le-v<0$ for $\phi_a\le\phi\le\phi_b$, with $V$ negative for
$\phi>\phi_c$ and $\phi_b$ large enough that $|V(\phi_b)|$ exceeds the energy available.
During expansion $\dot\rho=-3H\dot\phi^2\le0$, so $\rho$ is non-increasing and is bounded below by
zero. Since $\ddot\phi+3H\dot\phi=-V'\ge v$, the quantity $u=a^3\dot\phi$ obeys $\dot u\ge va^3$ and
therefore $\dot\phi\ge v\,a^{-3}\!\int a^3dt$. For a Hubble rate bounded by $H_{\max}$ the right-hand
side is bounded below by a positive constant $\delta=v/(3H_{\max})$ after a transient. Using
$H=\sqrt{\kappa\rho}$ for $\rho\ll\rho_c$,
$d\sqrt\rho/dt=-\tfrac32\sqrt\kappa\,\dot\phi^2\le-\tfrac32\sqrt\kappa\,\delta^2$,
so $\rho$ reaches zero in a time not exceeding $2\sqrt{\rho_i}/(3\sqrt\kappa\,\delta^2)$, and the
turnaround occurs. The recollapse is thus a property of the shape of the potential. The time at
which it occurs is not predicted by the construction; it is set by the length of the thawing
plateau, which is the quantity entering the entropy accounting of Section~\ref{sec:tolman}.

\section{Background dynamics}
\label{sec:bg}

We integrate $\dot\phi=\phi_v$, $\dot\phi_v=-3H\phi_v-V'(\phi)$, and Eq.~\eqref{eq:accel}
directly in cosmic time, with $\kappa=1$ units and the potential above, using an implicit stiff
(Radau) integrator with relative tolerance $10^{-10}$--$10^{-11}$, switching to smaller step caps
through the genuinely stiff dive-and-bounce transition, and monitoring the Friedmann-constraint
residual $|H^2-\rho(1-\rho/\rho_c)|/H^2$ throughout as an independent accuracy check (it remains
$\lesssim10^{-5}$ away from the immediate neighborhood of $H=0$ crossings, where the residual is
not a meaningful diagnostic since both sides vanish).

\begin{figure}[h]
\centering
\includegraphics[width=0.95\textwidth]{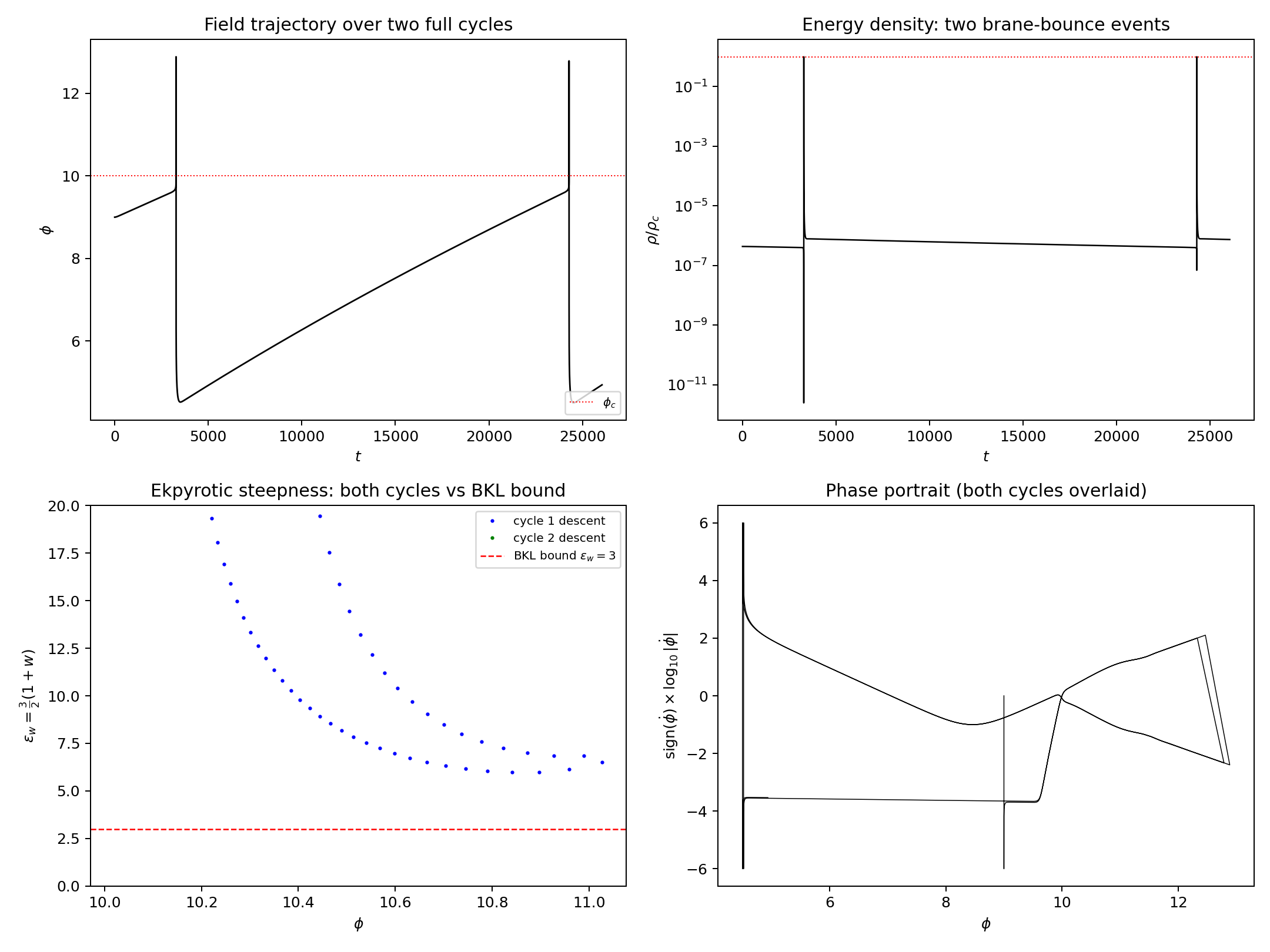}
\caption{Two complete cycles of the exact background evolution. \emph{Top left:} field
trajectory $\phi(t)$: a slow thaw to $\phi_c$ (dotted), rapid dive, bounce/wall complex, re-thaw,
twice. \emph{Top right:} $\rho/\rho_c$ on a log scale, touching the brane threshold at each
bounce. \emph{Bottom left:} the matter equation-of-state exponent $\epsilon_w=\frac32(1+w)$
during each ekpyrotic descent against the BKL bound $\epsilon_w=3$. \emph{Bottom right:} phase
portrait, both cycles overlaid, showing convergence toward a closed loop.}
\label{fig:twocycles}
\end{figure}

Figure~\ref{fig:twocycles} shows two complete cycles starting from the thawing plateau. $\phi$
climbs slowly until crossing the gate near $\phi_c=10$; $H$ passes smoothly through zero from
positive to negative with no external input. $\rho/\rho_c$ reaches $0.99996$ (cycle~1) and
$0.9988$ (cycle~2) at the two bounces, with $H$ passing smoothly from negative to positive and
$\ddot a>0$: a genuine non-singular bounce. Because the brane floor and the reflecting wall sit a
finite distance apart in field space, the field undergoes several rapid, damped
brane-bounce/wall-reflection oscillations before escaping back over the potential hump into the
slow-roll region, each oscillation shedding the $\rho\sim\rho_c$ energy it briefly carried to
Hubble friction and landing back on a low-energy ($\rho\sim10^{-7}\rho_c$) slow-roll trajectory.
The state after cycle~1 ($\phi=4.9307$, $\dot\phi=2.80\times10^{-4}$, $H=5.815\times10^{-3}$) and
after cycle~2 ($\phi=4.9348$, $\dot\phi=2.95\times10^{-4}$, $H=5.508\times10^{-3}$) agree to
$\sim0.1\%$ in $\phi$ and $\sim5\%$ in $H$: two dynamically independent passes through the
numerically delicate dive-bounce-escape sequence land the system back in nearly the same place,
numerical evidence of convergence toward a limit cycle. In non-inflationary cyclic
constructions generally, reaching such a limit cycle requires iterative tuning over successive
cycles \cite{IjjasSteinhardt2019NoBB}.

\begin{figure}[h]
\centering
\includegraphics[width=0.95\textwidth]{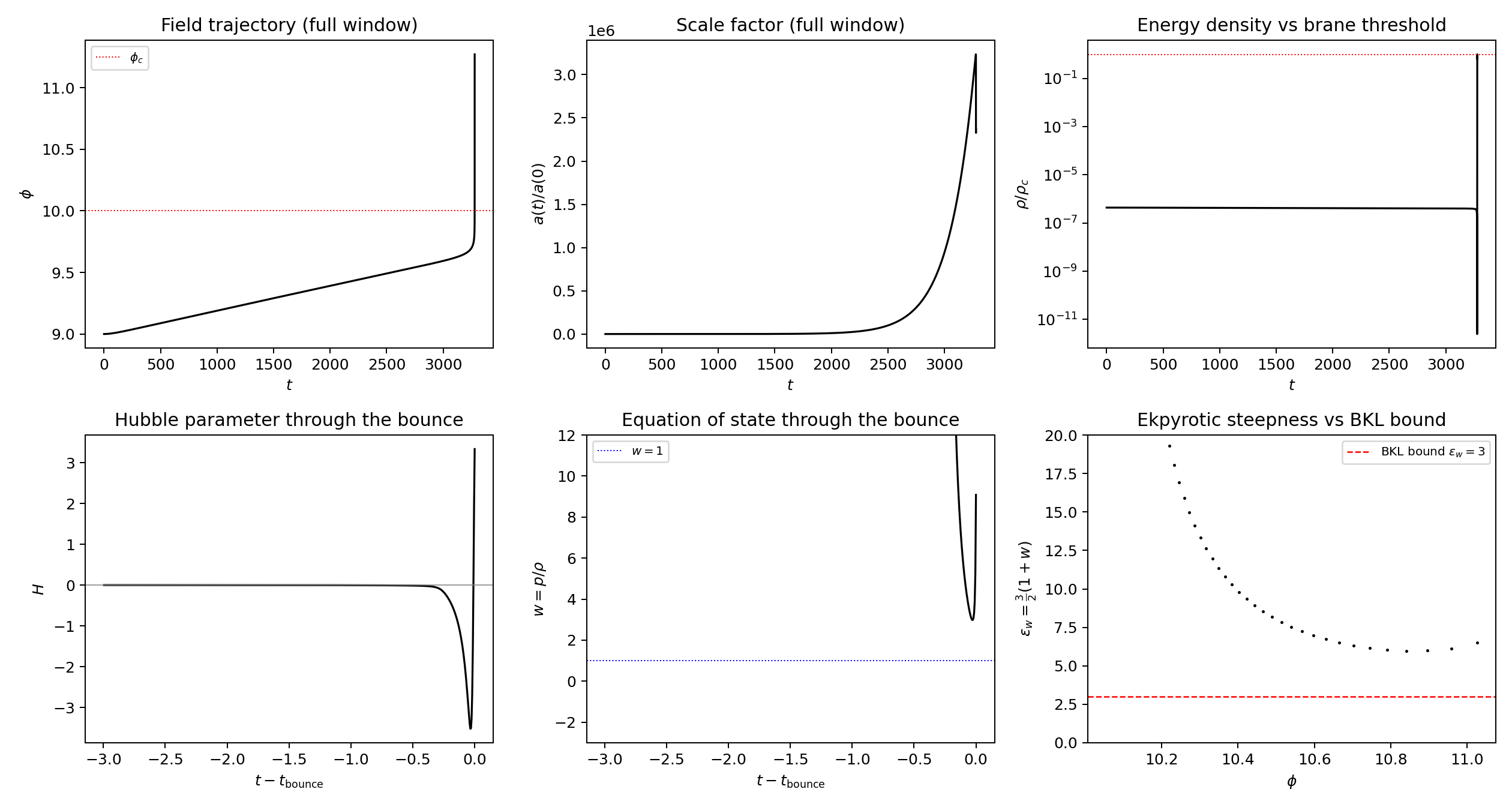}
\caption{Detail of the plateau$\to$descent$\to$ekpyrosis$\to$bounce sequence (cycle~1, first
bounce). \emph{Top row:} field, scale factor, and $\rho/\rho_c$ over the full window.
\emph{Bottom row:} Hubble parameter and equation of state through the bounce, and the
BKL-safety exponent $\epsilon_w(\phi)$ during the descent.}
\label{fig:bouncecycle}
\end{figure}

The relevant criterion, derived on the brane
in~\cite{ShtanovSahni2003} for the Bianchi~I generalization of Eq.~\eqref{eq:friedmann}, is that
the shear scalar $\sigma_{\alpha\beta}\sigma^{\alpha\beta}\propto a^{-6}$ must be overwhelmed by
the $\rho^2/M^6$ term, which scales as $a^{-2\cdot2\epsilon_w}$ for matter with
$\epsilon_w\equiv\frac32(1+w)$ (so $\rho\propto a^{-2\epsilon_w}$). This requires $\epsilon_w>3$
throughout the contraction, independent of the initial shear amplitude, exactly the mechanism
identified in the original ekpyrotic proposal~\cite{KhouryOvrutSteinhardtTurok2001} for solving
the BKL instability that plagues generic contracting cosmologies. The correct diagnostic is the
matter exponent $\epsilon_w=\frac32(1+w)$, computed from $w=p/\rho$, \emph{not} the geometric
$\epsilon_H\equiv-\dot H/H^2$: the two coincide only when the brane correction is negligible, and
$\epsilon_H$ is driven away from $\epsilon_w$, even changing sign, as $\rho\to\rho_c/2$. This is
precisely the brane term turning $\dot H$ around (Section~\ref{sec:brane_ns}), and has no bearing
on the anisotropy criterion. Table~\ref{tab:bkl} summarizes both cycles: $\epsilon_w$ stays
comfortably above the bound throughout, with a minimum safety margin of about a factor of two.

\begin{table}[h]
\centering
\begin{tabular}{lccc}
\toprule
& $\epsilon_w$ at onset ($\phi\to\phi_c^+$) & $\epsilon_w$ minimum & margin over bound \\
\midrule
Cycle 1 & $\sim99$ (transient at $V\to0$) & $5.97$ & $2.0\times$ \\
Cycle 2 & $\sim95$ (transient at $V\to0$) & $6.1\phantom{0}$ & $2.0\times$ \\
\bottomrule
\end{tabular}
\caption{BKL-safety exponent during the two ekpyrotic descents. The large transient values
immediately after the sign change are a coordinate artifact of dividing by $\rho\to0$ at the
zero-crossing, not a physical pathology, and are excluded from the quoted minima.}
\label{tab:bkl}
\end{table}

Because $\epsilon_w$ never approaches the bound from below, the model as constructed is free of
the BKL/chaotic-mixmaster instability and of the milder anisotropic-growth problem that afflicts
ekpyrotic models with $\epsilon_w$ too close to $3$.

\section{Tolman's entropy problem}
\label{sec:tolman}

\subsection{The classical argument and why spatial flatness evades it}

Tolman's 1934 objection to oscillating cosmology~\cite{Tolman1934} is a thermodynamic one. In a
\emph{closed} FRW universe with finite comoving volume $V_c$, entropy produced by dissipative
processes during one cycle (particle production, structure formation, black hole formation and
evaporation) adds to the total entropy $S$ contained in that finite volume. Because the total
energy in a closed universe returns to the same value at the same phase of successive cycles (by
the periodicity of the Friedmann equation with fixed curvature), while the entropy strictly
increases, the universe must occupy a larger comoving volume, and hence, since curvature fixes the
relation between comoving and physical volume in a closed universe, a larger physical volume, at
turnaround in each successive cycle to accommodate more entropy at fixed energy. Successive cycles
grow monotonically longer and larger; run backward in time, this implies a beginning a finite
number of cycles ago, not a truly past-eternal universe, and run forward, cycles without limit in
size and duration.

The construction of this paper is spatially flat, $\kappa_{\rm FRW}=0$ in
Eq.~\eqref{eq:friedmann}: comoving volume is infinite. There is no finite total entropy
$S=s\times V_c$ to bound, because $V_c=\infty$; Tolman's argument, which is a statement about a
finite quantity growing without bound, simply does not apply. This is not a technicality invented
to dodge the problem: it is the resolution given in the modern cyclic-universe
literature~\cite{SteinhardtTurok2002PRD,SteinhardtTurok2005,IjjasSteinhardt2019NoBB}, and it is
the reason those constructions, and this one, insist on exact spatial flatness rather than a
closed universe with curvature radius taken to be very large.

\subsection{Causal structure of an infinite cyclic universe}
\label{sec:horizon}

Spatial infinity removes Tolman's bound on the total entropy, but it raises two further questions.
One is whether the horizon problem becomes infinitely severe, and the other is whether the bounce
must occur simultaneously throughout an infinite space. Neither is required by the construction.
Equation~\eqref{eq:friedmann} is local: each region bounces when its own density reaches
$\rho_c$, and nothing in the model synchronizes distant regions. What must hold is that a comoving
region containing the observable universe, and a neighborhood of it larger than the present Hubble
radius, be causally connected and smooth at the onset of the ekpyrotic phase. This is the same
requirement as in any cosmology that does not assume global homogeneity, and it is the only
requirement that inflation itself addresses.

\emph{Causal contact.} Let $\Gamma\equiv a_{n+1}/a_n=e^{N_{\rm cyc}}$ be the ratio of scale factors
at the same phase of successive cycles, with $N_{\rm cyc}$ the net number of e-folds per cycle
defined in Section~\ref{sec:quant}. The scale factor is continuous through the non-singular
bounce, so comoving coordinates are common to all cycles. A comoving region equal to our present
Hubble volume then occupied a fraction $e^{-N_{\rm cyc}}$ of the comoving Hubble radius at the same
phase of the preceding cycle, which is $e^{-3N_{\rm cyc}}$ of its volume. For
$N_{\rm cyc}\gtrsim77$ this is a linear fraction below $10^{-33}$. The region was therefore
contained deep inside a single Hubble patch of the preceding cycle at the onset of its smoothing
phase, and the subsequent dark-energy phase and ekpyrotic contraction only decrease the comoving
Hubble radius. The same number of e-folds that dilutes the entropy in Section~\ref{sec:quant}
thus also supplies the separation of scales that the horizon problem requires, so the two
conditions are one.

\emph{Smoothness.} Contact does not by itself imply homogeneity. The smoothing is performed by the
contracting phase with $\epsilon_w>3$, which drives inhomogeneities and anisotropies toward zero
when they are initially smaller than the Hubble radius, as shown in numerical relativity
simulations with generic initial data \cite{Garfinkle2008,Erickson2004} and as is the basis of the
cyclic and bouncing proposals \cite{IjjasSteinhardt2018Inflation}. No statement is made here about
variations on scales larger than the Hubble radius at the onset of smoothing.

\emph{Non-simultaneity of the bounce.} In the separate-universe picture, neighboring regions
follow the same background trajectory shifted by $\delta N=\zeta$ e-folds of scale factor at fixed
density. With $\zeta_{\rm rms}=\sqrt{A_s}=4.6\times10^{-5}$ the corresponding spread in the time at
which regions reach a given density is of order $\zeta$ times the Hubble time of the contracting
phase, and, at the fiducial brane tension, of order $\zeta/H_{\max}$ with $H_{\max}=1.2\times10^7\,$GeV
in the bounce region. Regions inside the smoothed patch therefore bounce together to a relative
accuracy of order $10^{-4}$ of the bounce duration. Regions separated by more than the smoothed
patch need not bounce at the same time, and the model makes no prediction for them. The resulting
patchwork of regions is a feature of every non-inflationary and most inflationary completions of
an infinite universe, and the observable universe lies inside one smooth patch.

\subsection{Quantitative formulation of the problem}
\label{sec:quant}

Spatial flatness defeats the letter of Tolman's argument but not its spirit: an observer confined
to one Hubble patch, at the same phase of successive cycles, should not see a secularly increasing
entropy \emph{density}, or the model would predict an ever-messier universe unlike the one we
observe. The resolution, present already in the original cyclic proposal
\cite{SteinhardtTurok2002PRD} and made central to the requirement that the present accelerating
epoch be a permanent feature of the cycle in \cite{SteinhardtTurok2005}, is dilution: entropy
produced in comoving volume during one cycle is diluted, as physical volume expands, by the
\emph{following} dark-energy phase, before the next ekpyrotic contraction begins. Entropy in a
fixed comoving volume is (to the accuracy that dissipative production has ceased) conserved, so
entropy \emph{density} dilutes as $a^{-3}$; over $N$ e-folds of expansion, the density falls by $e^{-3N}$.

We make this quantitative with a definition that is independent of the time dependence of the
Hubble volume. Fix once and for all a comoving region $V_{\rm ref}$ whose physical volume at a
reference epoch (today) equals the Hubble volume at that epoch. Because $V_{\rm ref}$ is a fixed
comoving region, the divergence of the Hubble radius at $H=0$ plays no role. Let $s_n$ denote the
entropy density at the onset of the ekpyrotic phase of cycle $n$, and let $\Delta s$ be the entropy
density produced during a cycle, measured at the same phase. Entropy in the comoving region is
conserved after production, so entropy density dilutes as $a^{-3}$, and
\begin{equation}
s_{n+1}=\frac{s_n+\Delta s}{\Gamma^3},\qquad \Gamma=e^{N_{\rm cyc}},\qquad
s_\ast=\frac{\Delta s}{\Gamma^3-1}.
\label{eq:recursion}
\end{equation}
The recursion converges geometrically, with ratio $\Gamma^{-3}$, to the fixed point $s_\ast$ for any
$N_{\rm cyc}>0$. The entropy density at the onset of ekpyrosis therefore does not grow secularly
as soon as the universe grows from cycle to cycle, and Tolman's growth of the cycles is absent.
The size of $s_\ast$ is what matters physically. Let $S_\star\equiv\Delta s\,a^3V_{\rm ref}$ be the
entropy produced in $V_{\rm ref}$ during a cycle. This is dominated, for our universe, not by the
cosmic microwave background ($S_{\rm CMB}\sim10^{88}\,k_B$) but by supermassive black holes formed
during structure formation, estimated at $S_{\rm BH}\sim10^{100}$--$10^{104}\,k_B$ depending on
assumptions about their ultimate accretion history~\cite{EganLineweaver2010}. Requiring that the
residual entropy in $V_{\rm ref}$ at the onset of the next ekpyrotic phase be below one unit of
$k_B$, which is a conservative criterion of negligibility, gives $\Gamma^3\gtrsim S_\star$, that is
\begin{equation}
N_{\rm cyc}\gtrsim\frac13\ln S_\star\approx76.8\ \ (S_\star=10^{100}),\qquad
N_{\rm cyc}\gtrsim\frac13\ln S_\star\approx79.8\ \ (S_\star=10^{104}).
\label{eq:tolmanbound}
\end{equation}
The quantity that enters is the net number of e-folds per cycle,
\begin{equation}
N_{\rm cyc}=N_{\rm exp}-N_{\rm contr}+N_{\rm DE},
\label{eq:Ncyc}
\end{equation}
where $N_{\rm contr}$ is the e-folds lost during the ekpyrotic contraction, $N_{\rm exp}$ is the
e-folds gained during expansion between the bounce and the onset of the next dark-energy phase,
and $N_{\rm DE}$ is the e-folds of accelerated expansion. Between two densities $\rho_1<\rho_2$
a fluid with $\rho\propto a^{-2\epsilon}$ changes the scale factor by
$\ln(\rho_2/\rho_1)/(2\epsilon)$ e-folds. With $\rho_c^{1/4}=10^{13}\,$GeV and the onset of
ekpyrosis at the dark-energy density, $\rho_{\rm on}\simeq\rho_\Lambda=(2.3\times10^{-3}\,{\rm eV})^4$,
one has $L\equiv\ln(\rho_c/\rho_{\rm on})\simeq227$. For the contraction with the running
steepness of Eq.~\eqref{eq:potential}, $\epsilon_w=c(\phi)^2/2$ with $c$ relaxing from $4.2$ to
$3.2$, and $d\ln\rho/d\phi\simeq c$, so
\begin{equation}
N_{\rm contr}=\int_0^{L}\frac{ds}{c(\phi(s))^2}\simeq21.6,\qquad d\phi/ds=1/c,
\end{equation}
which spans a field range $\Delta\phi\simeq70$ at the physical hierarchy. The expansion between
the bounce and the dark-energy density supplies $N_{\rm exp}=L/(2\epsilon_{\rm exp})$, where
$\epsilon_{\rm exp}=3$ for kination, the only post-bounce phase present in the model, and
$\epsilon_{\rm exp}=2$ if the post-bounce universe is radiation dominated. Since BKL safety
requires $\epsilon_c>3\ge\epsilon_{\rm exp}$, the difference $(L/2)(1/\epsilon_{\rm exp}-1/\epsilon_c)$
is positive: ekpyrotic contraction shrinks the universe less than the following expansion
enlarges it, so that the net growth per cycle is positive even before dark energy is included.
Table~\ref{tab:cycle} gives the resulting requirement on the dark-energy phase.

\begin{table}[h]
\centering
\begin{tabular}{lcccc}
\toprule
post-bounce expansion & $N_{\rm exp}$ & $N_{\rm exp}-N_{\rm contr}$ & $N_{\rm DE}$ ($S_\star=10^{100}$) & $N_{\rm DE}$ ($S_\star=10^{104}$) \\
\midrule
kination, $\epsilon_{\rm exp}=3$   & $37.8$ & $16.2$ & $60.5$ & $63.6$ \\
radiation, $\epsilon_{\rm exp}=2$ & $56.7$ & $35.1$ & $41.6$ & $44.7$ \\
\bottomrule
\end{tabular}
\caption{Net e-folds per cycle and the resulting requirement on the dark-energy phase for
$L=227$ and $N_{\rm contr}=21.6$. The first row is the post-bounce history of the explicit model,
the second row applies if radiation domination follows the bounce.}
\label{tab:cycle}
\end{table}

The dark-energy phase must therefore supply of order $N_{\rm DE}\sim42$--$64$ e-folds, depending on
the post-bounce history, and the total $N_{\rm cyc}\sim77$--$80$ is fixed by the entropy. That is
a large number, but a finite and unremarkable one on cosmological timescales, not a fine-tuning
requirement in the way the number $10^{100}$ itself might suggest. If the present accelerating
epoch persists at its current rate for a further time $\Delta t$, it supplies
$N_{\rm DE}\approx H_0\Delta t$ e-folds; using $H_0^{-1}\approx$ the current age of the universe
(order-of-magnitude, for a de Sitter-like equation of state), $N_{\rm DE}\sim60$ requires only
$\Delta t\sim60$ times the current age of the universe. That is a long time on a human scale, but
a vanishingly short one compared to the eternity a truly cyclic model has available, and far shorter
than the time such models generally posit between the present epoch and the eventual turnaround
\cite{SteinhardtTurok2005,IjjasSteinhardt2019NoBB}. The criterion of one unit of $k_B$ per
reference volume is conservative, and convergence of the recursion, Eq.~\eqref{eq:recursion}, itself
requires only $N_{\rm cyc}>0$.

\subsection{Status of this requirement in the explicit model}

The background integration of Section~\ref{sec:bg} uses a dark-energy plateau lasting a
deliberately short, sub-e-folding stretch of expansion, compressed for numerical tractability in
exactly the same way, and for the same reason, that the energy hierarchy between $V_\Lambda$,
$V_0$, and $\rho_c$ is compressed there (Section~\ref{sec:discussion}). Nothing about the
mechanism of Section~\ref{sec:matter}, a thawing scalar field descending toward a sign-changing
gate, prevents the plateau from lasting the required $N_{\rm DE}\sim42$--$64$ e-folds instead:
extending $\phi_0$ and the distance in field space between the plateau's slow-roll region and the
gate at $\phi_c$ lengthens the plateau phase without altering the turnaround mechanism of
Section~\ref{sec:turnaround}, since that mechanism depends only on the field crossing zero energy
density with nonzero kinetic energy, and the proof of inevitability given there does not depend on
how long the preceding slow roll lasted. The accounting of Eq.~\eqref{eq:Ncyc} is carried out at
the physical hierarchy, $L=227$, whereas the integrations of Section~\ref{sec:bg} use a compressed
one. The dimensionless quantities that control the background, $\epsilon_w$ and $x=\rho/\rho_c$,
do not depend on the absolute scale, and the compression changes $N_{\rm contr}$ and
$N_{\rm exp}$ only through $L$ in the formulas above. The background has not been re-integrated at the physical
duration, but the number required is of the kind these constructions already contemplate rather
than an obstruction specific to the braneworld realization presented here.

\section{Perturbation theory: the single-field blue-tilt problem}
\label{sec:pert}

\subsection{Mukhanov--Sasaki equation and exact constant-$\epsilon$ solution}

The comoving curvature perturbation obeys, in conformal time ($d\tau=dt/a$),
\begin{equation}
u_k''+\Big(k^2-\frac{z''}{z}\Big)u_k=0, \qquad \zeta_k=\frac{u_k}{z}, \qquad z\equiv a\sqrt{2\epsilon},
\label{eq:MS}
\end{equation}
with $\epsilon\equiv-\dot H/H^2$ the exact background slow-roll parameter, and $\zeta_k$ the Fourier mode of the comoving curvature perturbation $\mathcal R=\Phi+H\delta\phi/\dot\phi$ (Newtonian gauge), with $u_k=z\zeta_k$ its canonical variable. The applicability of this equation on the brane is established in Section~\ref{sec:brane_pert}: for CMB-scale modes, which exit the horizon many e-folds before the bounce, deep in $\rho\ll\rho_c$ where $\epsilon_H=\epsilon_w$ to high accuracy, the corrections are of order $\epsilon_w\rho/\rho_c$. For constant
$\epsilon>3$ the ekpyrotic attractor is $a(t)\propto(-t)^{1/\epsilon}$, which in conformal time
becomes $a(\tau)\propto(-\tau)^p$ with
\begin{equation}
p = \frac1{\epsilon-1},
\end{equation}
obtained by direct integration: $d\tau=dt/a\propto(-t)^{-1/\epsilon}dt$, so
$-\tau\propto(-t)^{1-1/\epsilon}=(-t)^{(\epsilon-1)/\epsilon}$, hence
$(-t)\propto(-\tau)^{\epsilon/(\epsilon-1)}$ and $a=(-t)^{1/\epsilon}\propto(-\tau)^{1/(\epsilon-1)}$.
Since $\epsilon$ is constant, $z=a\sqrt{2\epsilon}\propto a$, so $z''/z=a''/a$; for
$a\propto(-\tau)^p$, direct differentiation gives $a''/a=p(p-1)/\tau^2$, which we write as
$(\nu^2-\frac14)/\tau^2$ with
\begin{equation}
\nu = \Big|p-\frac12\Big|,
\end{equation}
the standard de-Sitter-like Hankel-function problem. Matching the sub-horizon mode to the
flat-space positive-frequency solution $u_k\to e^{-ik\tau}/\sqrt{2k}$ as $-k\tau\to\infty$, and
reading off the $k$-dependence of the super-horizon ($-k\tau\to0$) amplitude of $\zeta_k=u_k/z$
using the small-argument behavior of the Hankel function,
$H_\nu^{(1)}(x)\to-\frac{i}\pi\Gamma(\nu)(2/x)^\nu$, gives $|\zeta_k|^2\propto k^{-2\nu}$ and hence
$n_s-1=3-2\nu$, i.e.
\begin{equation}
n_s = 4-2\Big|\frac1{\epsilon-1}-\frac12\Big|.
\label{eq:nsconst}
\end{equation}
For $\epsilon\gg1$ this gives $n_s\to3$, the well-known ``badly blue'' single-field ekpyrotic
result~\cite{Lyth2002}. We solved Eq.~\eqref{eq:MS} numerically (complex ODE integration from deep
sub-horizon, $|k\tau|\gg1$, to deep super-horizon, $|k\tau|\ll1$, for a range of $k$, followed by a
power-law fit of $k^3|\zeta_k|^2$ vs.\ $k$) for $\epsilon=12$ and found
$n_s^{\rm numerical}=3.18176$ against $n_s^{\rm analytic}=3.18182$ from Eq.~\eqref{eq:nsconst},
agreement to $6\times10^{-5}$. This validates both the mode-equation solver and the numerical
spectral-index extraction procedure used throughout this paper.

\subsection{The model's actual running does not fix it}

We solved Eq.~\eqref{eq:MS} on a semi-analytic attractor background with $\epsilon(\tau)$ running
smoothly (logarithmically in $-\tau$, i.e.\ uniformly in e-folds) from $\epsilon_i=19$ (the onset
value found in the explicit potential) down to $\epsilon_f=6$ (the asymptotic value the running
settles toward), constructed by integrating the exact relations
$\mathcal H'=\mathcal H^2(1-\epsilon(\tau))$, $(\ln a)'=\mathcal H$ ($\mathcal H\equiv a'/a=aH$)
directly, without assuming the constant-$\epsilon$ attractor form, so the calculation
captures the true effect of the running on $z''/z$, not merely a slow-roll-order estimate.

\begin{figure}[h]
\centering
\includegraphics[width=0.55\textwidth]{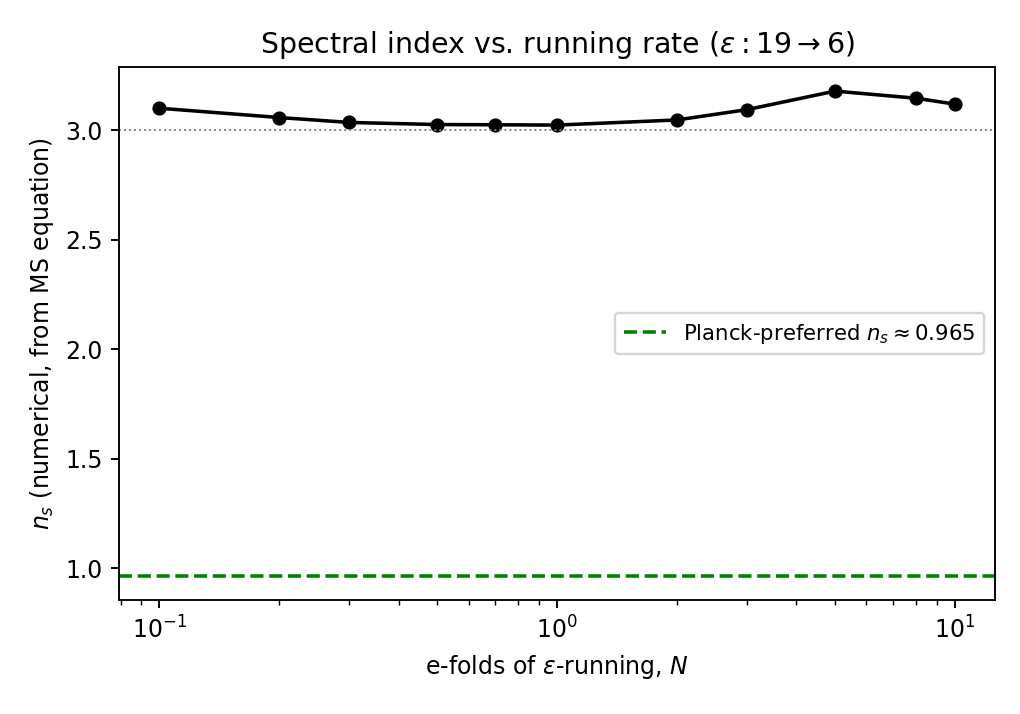}
\caption{Spectral index vs.\ the rate at which $\epsilon$ runs from $19$ to $6$ (fixed endpoints).
The result is pinned near $n_s\approx3$ for any running rate tested, far from the observed
$n_s\approx0.965$.}
\label{fig:nsrun}
\end{figure}

Figure~\ref{fig:nsrun} shows the result is essentially flat and pinned near $n_s\approx3$
regardless of the running rate. The reason a mild running does not help is visible already in
Eq.~\eqref{eq:nsconst}: no constant $\epsilon>3$ gives $\nu=3/2$ (exact scale invariance would
need $\epsilon<3/2$, excluded outright by the BKL bound $\epsilon_w>3$), so rescuing $n_s$
requires the time-dependence of $\epsilon$ itself to make an $O(1)$, not perturbative,
contribution to $z''/z$, sustained over the observationally relevant window. A BKL-safe,
monotonic running of the kind produced naturally by Eq.~\eqref{eq:potential} does not supply
this. This is not a numerical failure but the structural reason the ekpyrotic literature moved,
in two distinct directions, toward either a substantially more specialized designer running
$\epsilon(N)$ than the one used here (the adiabatic-ekpyrosis
construction~\cite{KhouryySteinhardt2010,LiSteinhardt2010}, not attempted in this paper) or the established entropic mechanism of a second field
\cite{Notari2002,Finelli2002,LehnersEtAl2007}, which we develop fully in
Section~\ref{sec:entropic}.

\subsection{Perturbation equations on the brane}
\label{sec:brane_pert}

The mode equations used so far are those of four-dimensional general relativity. On the brane the
induced Einstein equations contain, besides the stress-energy of the brane matter, a term quadratic
in the stress-energy, whose homogeneous limit gives the factor $1-\rho/\rho_c$ in
Eq.~\eqref{eq:friedmann}, and the projection $E_{ab}$ of the bulk Weyl tensor, which acts as a
traceless fluid with energy density $\delta\rho_E$, momentum potential $\delta q_E$, and
anisotropic stress $\delta\pi_E$ \cite{ShiromizuMaedaSasaki2000,Maartens2004}. Scalar perturbations
of the closely related bouncing brane were analyzed in Ref.~\cite{Maier2013}, which obtained the
perturbed brane equations for dust and radiation on a closed brane and showed that the
perturbations stay bounded through the bounce. We adapt that system to the present flat brane with
a scalar field, and we use it to establish where the four-dimensional equations of
Sections~\ref{sec:pert}--\ref{sec:brane_ns} can be trusted. In this subsection $8\pi G=1$ and
$x\equiv\rho/\rho_c$.

\emph{Effective fluid.} Define the effective density and pressure by
$\rho_{\rm eff}=\rho(1-x)$ and $p_{\rm eff}=p-x(2p+\rho)$, so that
$H^2=\rho_{\rm eff}/3$ and $\dot H=-(\rho_{\rm eff}+p_{\rm eff})/2$, which is
Eq.~\eqref{eq:effnec}. Perturbing in the Newtonian gauge,
$ds^2=-(1+2\Phi)dt^2+a^2(1-2\Psi)d\mathbf x^2$, the corresponding effective perturbations are
\begin{equation}
\delta\rho_{\rm eff}=(1-2x)\delta\rho,\qquad
\delta q_{\rm eff}=(1-2x)\delta q,\qquad
\delta p_{\rm eff}=(1-2x)\delta p-2x(1+w)\,\delta\rho,
\label{eq:effpert}
\end{equation}
with $w=p/\rho$. For a scalar field $\delta q=-\dot\phi\,\delta\phi$ and
$\delta\rho=\dot\phi\,\delta\dot\phi-\dot\phi^2\Phi+V'\delta\phi$.

\emph{Closed system.} With $\delta\pi_E=0$, so that $\Phi=\Psi$ as in Ref.~\cite{Maier2013}, the
Einstein equations and the conservation of the total source give
\begin{align}
-\frac{k^2}{a^2}\Phi-3H(\dot\Phi+H\Phi)&=\tfrac12\big[(1-2x)\delta\rho-\delta\rho_E\big],
\label{eq:b00}\\
\dot\Phi+H\Phi&=\tfrac12\big[(1-2x)\dot\phi\,\delta\phi+\delta q_E\big],
\label{eq:b0i}\\
\delta\dot\rho_E+4H\delta\rho_E&=\frac{k^2}{a^2}\delta q_E,
\label{eq:bE1}\\
\delta\dot q_E+3H\delta q_E+\tfrac13\delta\rho_E&=-2x(1+w)\big(\delta\rho+3H\dot\phi\,\delta\phi\big),
\label{eq:bE2}
\end{align}
together with the scalar-field equation
$\delta\ddot\phi+3H\delta\dot\phi+(k^2/a^2+V'')\delta\phi=4\dot\phi\dot\Phi-2V'\Phi$, which is exact
on the brane because the matter is conserved. Equations~\eqref{eq:bE1}--\eqref{eq:bE2} are the
equations of the Weyl fluid of Ref.~\cite{Maier2013} for a flat brane, written with our sign
convention for the momentum potentials. They are derived in Appendix~\ref{app:brane} from the
requirement that the matter, the quadratic correction, and the Weyl fluid together be conserved. The
energy equation of the effective fluid is satisfied identically, so $\delta\rho_E$ is sourced
only through its coupling to $\delta q_E$, and the Weyl fluid is driven by the matter only through the
momentum equation, with the factor $2x(1+w)=2\dot\phi^2/\rho_c$. The system is therefore driven by
$x$ and vanishes smoothly in the limit $\rho/\rho_c\to0$.

A point about the bulk follows. For fluids without anisotropic stress the ansatz of an unperturbed
de Sitter bulk, $\delta E_{ab}=0$, forces the metric perturbation to vanish \cite{Maier2013}, so
the perturbations of a scalar field necessarily excite the Weyl fluid, and the system above, which
retains $\delta\rho_E$ and $\delta q_E$, is the minimal consistent one. The closure $\delta\pi_E=0$
is an assumption on the bulk. It is the one used in Ref.~\cite{Maier2013}, it is appropriate on
large scales, and it is the weakest point of any brane-based perturbation theory of this type, so we
state where our conclusions do and do not depend on it.

\emph{Numerical test of the system and size of the brane corrections.} We solved
Eqs.~\eqref{eq:b0i}--\eqref{eq:bE2} and the scalar-field equation for modes on the
exponential-potential contracting attractor with $\epsilon_w=c^2/2=8$, on the exact brane
background, starting each mode deep inside the horizon in the Bunch--Davies state of the
four-dimensional equations at a density fraction below $10^{-8}$ where the corrections are
negligible. The constraint~\eqref{eq:b00}, which is not used in the evolution, is preserved by the
evolution to a relative accuracy between $10^{-15}$ and $2\times10^{-9}$ in all cases of
Table~\ref{tab:bpert}, and it is violated at the level of $7\times10^{-6}$ to $4\times10^{-4}$ if the Weyl fluid is
omitted. This confirms that the equations are mutually consistent. Table~\ref{tab:bpert} compares the
spectral index of $k^3|\mathcal R_k|^2$ for modes that leave the horizon at a density fraction
$x_\ast$, evaluated when the field reaches the value at which $\rho/\rho_c=0.05$ on the brane
background, with the same procedure applied to the four-dimensional equations on the corresponding
four-dimensional background. For constant $\epsilon_w=8$ the asymptotic value of
Eq.~\eqref{eq:nsconst} is $n_s=3.286$.

\begin{table}[h]
\centering
\begin{tabular}{ccccccc}
\toprule
$x_\ast$ & $n_s$ (brane) & $n_s$ (GR) & $\Delta n_s$ & $P_{\rm brane}/P_{\rm GR}$ & constraint & $|\mathcal R|_{\rm no\,Weyl}/|\mathcal R|$ \\
\midrule
$10^{-5}$ & $3.2766$ & $3.2757$ & $9.5\times10^{-4}$ & $1.0013$ & $1\times10^{-15}$ & \\
$10^{-4}$ & $3.2676$ & $3.2655$ & $2.0\times10^{-3}$ & $1.0028$ & $4\times10^{-14}$ & $0.9984$ \\
$10^{-3}$ & $3.2508$ & $3.2461$ & $4.7\times10^{-3}$ & $1.0059$ & $9\times10^{-12}$ & \\
$10^{-2}$ & $3.2237$ & $3.2115$ & $1.2\times10^{-2}$ & $1.0137$ & $2\times10^{-9}$ & $0.9923$ \\
\bottomrule
\end{tabular}
\caption{Spectral index of the adiabatic mode computed from the closed brane system,
Eqs.~\eqref{eq:b0i}--\eqref{eq:bE2}, and from four-dimensional general relativity, for modes leaving
the horizon at density fraction $x_\ast$. The last column compares the curvature perturbation with
and without the Weyl fluid. The residual shift $\Delta n_s$ at the smallest $x_\ast$ is set by the
finite value of $k/aH$ at the common evaluation point and by the evolution to
$\rho/\rho_c=0.05$, and it does not scale with $x_\ast$.}
\label{tab:bpert}
\end{table}

The brane corrections shift the tilt by at most $10^{-2}$ for $x_\ast\le10^{-2}$, and the
power spectrum changes by at most $1.4\%$. The spectral index remains at
$3.2$--$3.3$, as in general relativity. The Weyl fluid changes the curvature perturbation by
$0.16\%$ at $x_\ast=10^{-4}$ and $0.8\%$ at $x_\ast=10^{-2}$, consistent with a correction of order
$\epsilon_wx_\ast$. For the largest observable mode, which leaves the horizon at
$x_\ast\sim10^{-11}$, any correction attributable to the brane is smaller by many orders of
magnitude. The four-dimensional mode equation is therefore accurate for the modes that are
compared with observation, and this conclusion does not depend on the closure $\delta\pi_E=0$
except through terms of order $x_\ast$.

\begin{figure}[h]
\centering
\includegraphics[width=0.85\textwidth]{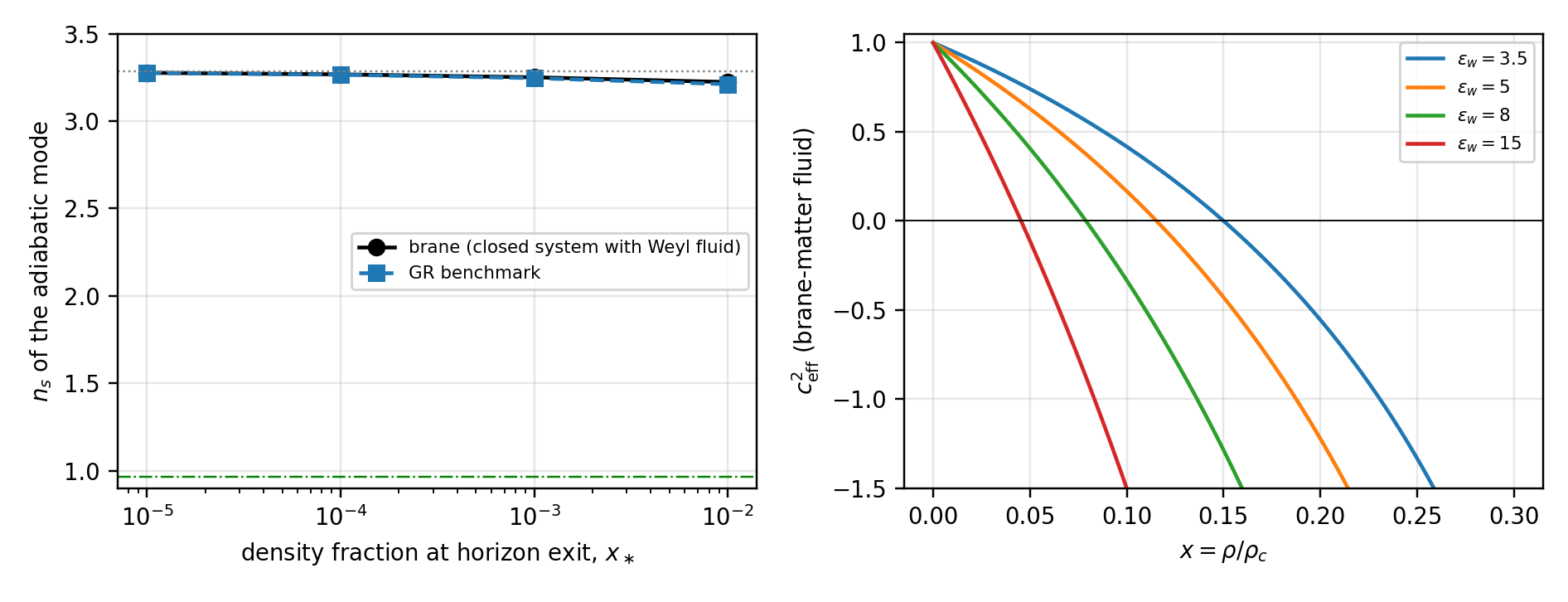}
\caption{Left: spectral index of the adiabatic mode from the closed brane system and from
four-dimensional general relativity as a function of the density fraction at horizon exit. The
dotted line is the constant-$\epsilon$ value $3.286$ and the dash-dotted line is the observed
$0.9649$. Right: sound speed of the effective brane-matter fluid, Eq.~\eqref{eq:ceff}, which
vanishes at $x_0=1/(2+4\epsilon_w/3)$.}
\label{fig:bpert}
\end{figure}

\emph{Limits of the four-dimensional equation.} Two features restrict the range of applicability
of Eq.~\eqref{eq:MS}. First, $z=a\dot\phi\sqrt{1-2x}/H$ diverges at $H=0$ and vanishes at
$x=\tfrac12$, so the Mukhanov--Sasaki variable cannot be continued through either point, and
the matching across the bounce has to be carried out for a variable that is regular there. Second,
the rest-frame sound speed of the effective fluid of Eq.~\eqref{eq:effpert}, obtained from
$\delta p_{\rm eff}/\delta\rho_{\rm eff}$ in the gauge $\delta\phi=0$ where $\delta p=\delta\rho$, is
\begin{equation}
c_{\rm eff}^2=1-\frac{4\epsilon_w}{3}\,\frac{x}{1-2x},
\label{eq:ceff}
\end{equation}
which vanishes at $x_0=1/(2+4\epsilon_w/3)$. This is $0.15$ for $\epsilon_w=3.5$, $0.079$ for
$\epsilon_w=8$, and $0.024$ for $\epsilon_w=30$ (Figure~\ref{fig:bpert}, right). Beyond $x_0$ the
brane matter fluid has a gradient instability unless the bulk contribution compensates it, and the
vacuum state of short-wavelength modes cannot be specified from brane data. Modes that cross the
horizon at $x>x_0$ are therefore outside the domain in which a prediction can be made from the
brane equations alone.

\emph{The entropy sector.} The kinetically coupled field of Section~\ref{sec:ridge} is a spectator,
$\bar\chi=0$, so its stress-energy is quadratic in $\delta\chi$ and does not source the scalar
metric perturbations at linear order. Its quadratic action on the brane background,
$S^{(2)}_\chi$ of Section~\ref{sec:ridge}, involves only the induced metric and the background
$a(t)$ and $\phi(t)$, which are given exactly by Eq.~\eqref{eq:friedmann}, and the matter is
confined to the brane. The canonical variable $u=a\sqrt{f}\,\delta\chi$ and its mode equation are
therefore exact on the brane, with no dependence on the bulk, and the entropy-sector tilt of
Sections~\ref{sec:ridge}--\ref{sec:running} does not rely on the closure $\delta\pi_E=0$. The bulk
enters only through the matching of $\delta\chi$ to $\zeta$ at the wall and through the growth of
$\delta\chi$ near the bounce, treated in Sections~\ref{sec:conversion} and \ref{sec:kineticstability}.

\section{Brane corrections cannot fix $n_s$}
\label{sec:brane_ns}

\subsection{The brane term drives $\epsilon_H$ away from $\epsilon_w$}

\begin{figure}[h]
\centering
\includegraphics[width=0.95\textwidth]{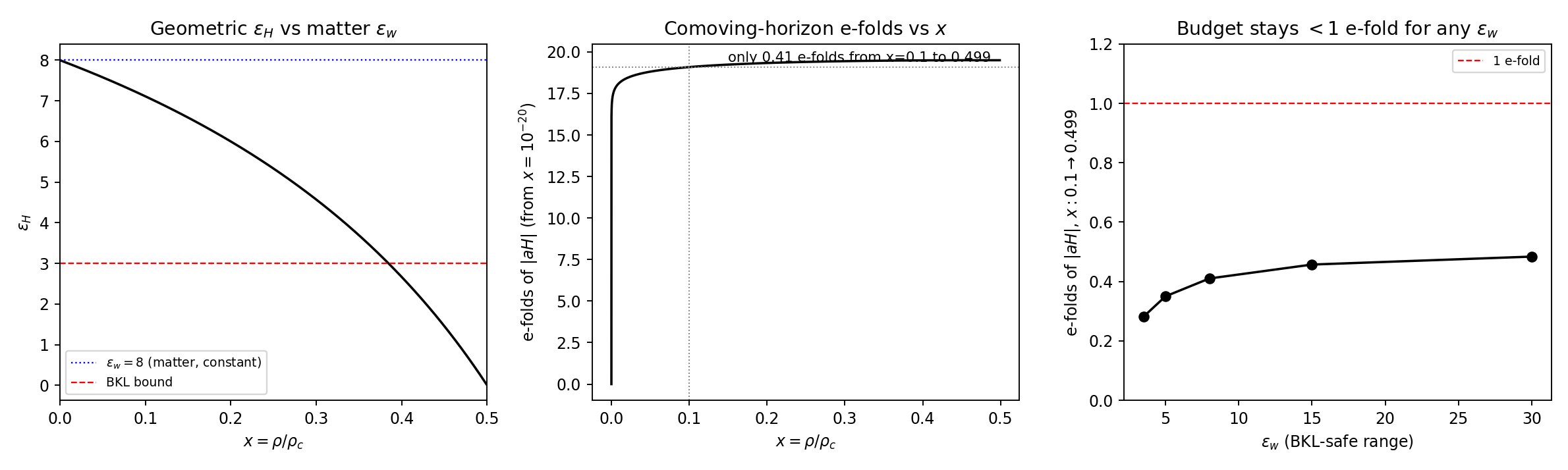}
\caption{\emph{Left:} exact geometric $\epsilon_H(x)$ (Eq.~\eqref{eq:epsH}) for fixed
$\epsilon_w=8$, compared to the constant $\epsilon_w$ and the BKL bound. \emph{Center:} e-folds
of comoving-horizon shrinkage $|aH|$ vs.\ $x=\rho/\rho_c$: essentially all $19.5$ e-folds
available down to $x=10^{-20}$ are used up before $x=0.1$, and only $0.41$ further e-folds remain
between $x=0.1$ and the domain edge $x=0.499$. \emph{Right:} this $<1$-e-fold budget persists
across the entire BKL-safe range of $\epsilon_w$.}
\label{fig:branebudget}
\end{figure}

Section~\ref{sec:pert} worked in the regime $\rho\ll\rho_c$, where $\epsilon_H=\epsilon_w$ to high
accuracy. Since the brane term modifies the background $H(t)$ nonlinearly, the natural next
question is whether it also modifies the \emph{geometric} $\epsilon_H$ that actually enters
Eq.~\eqref{eq:MS} in a way that produces a realistic spectrum once modes are allowed to cross the
horizon closer to the bounce, where $\rho/\rho_c$ is no longer negligible. From
Eqs.~\eqref{eq:friedmann}--\eqref{eq:accel}, using $\epsilon_H=-\dot H/H^2$ directly,
\begin{equation}
\epsilon_H = \epsilon_w\,\frac{1-2x}{1-x}, \qquad x\equiv\frac\rho{\rho_c},
\label{eq:epsH}
\end{equation}
exact, with no attractor or slow-roll approximation, and reducing to $\epsilon_H=\epsilon_w$ as
$x\to0$ as required. Even holding the matter exponent fixed at a comfortably BKL-safe value
$\epsilon_w=8$, Eq.~\eqref{eq:epsH} shows $\epsilon_H$ driven purely by the brane nonlinearity
from $8$ at $x=0$ down to $0$ at $x=\frac12$, passing through $\epsilon_H\approx1.5$, the value
that, if sustained, would give exact scale invariance ($\nu=3/2$) in Eq.~\eqref{eq:nsconst},
around $x\approx0.43$ (Figure~\ref{fig:branebudget}, left).

\subsection{The e-fold budget for this transition is structurally too small}
\label{sec:budget72}

The relevant question is not whether $\epsilon_H$ passes through a nice value, but over how many
e-folds of comoving-horizon shrinkage, $N\equiv\ln|aH|$, it does so, since that determines how
many decades of $k$ could be affected. This is computed directly, with no approximation: matter
conservation at fixed $\epsilon_w$ gives $a(x)=(x/x_i)^{-1/(2\epsilon_w)}$ exactly, and
Eq.~\eqref{eq:friedmann} gives $H(x)=-\sqrt{x(1-x)}$ exactly (units $\rho_c=1$). The budget from
$x=0.1$ to $x=0.499$ is only $N\approx0.41$ for $\epsilon_w=8$
(Figure~\ref{fig:branebudget}, center), and repeating for $\epsilon_w\in\{3.5,5,8,15,30\}$, the
entire BKL-safe range, the budget stays below one e-fold throughout
(Figure~\ref{fig:branebudget}, right). This is because $a(x)$ is almost exactly constant near
$x\sim0.3$--$0.5$ for any BKL-safe $\epsilon_w$ (its logarithmic rate of change is
$-1/(2\epsilon_w)\lesssim-0.14$ per decade of $x$, and the available range in $x$ is itself less
than one decade), so essentially all of the comoving-horizon shrinkage is generated by the bounded
function $H(x)=\sqrt{x(1-x)}$ alone, which cannot vary by more than an $O(1)$ factor between
$x=0$ and $x=\frac12$. This is a structural, parameter-independent property of the simplest
($m=0$) Shtanov--Sahni brane, not a matter of insufficient tuning of $\epsilon_w$ or $\rho_c$ (the
latter drops out of Eq.~\eqref{eq:epsH} entirely and only rescales the overall magnitude of $H$, not
the shape of $\epsilon_H(x)$ or the e-fold budget). 

This is a statement about the background. Whether modes that cross the horizon in the corresponding
interval are described by Eq.~\eqref{eq:MS} is a separate question, and it is where the structure
of the brane enters. Two features limit the range of Eq.~\eqref{eq:MS} (Section~\ref{sec:brane_pert}).
The variable $z=a\sqrt{2\epsilon_H}$ vanishes at $x=\tfrac12$ and diverges at $H=0$, so the
Mukhanov--Sasaki variable cannot be continued through those points. And the rest-frame sound
speed of the effective brane-matter fluid, Eq.~\eqref{eq:ceff}, vanishes at
$x_0=1/(2+4\epsilon_w/3)$, which for $\epsilon_w=8$ is $0.079$ and for the smallest BKL-safe value
$\epsilon_w=3.5$ is $0.15$. Beyond $x_0$ the effective fluid has a gradient instability unless a
bulk contribution removes it, and the initial state of short-wavelength modes is not determined by
brane data. The window $x:0.15\to0.45$ in which $\epsilon_H$ passes through $1.5$ for $\epsilon_w=8$
lies entirely beyond $x_0$, so a spectral index computed from Eq.~\eqref{eq:MS} there is not a
prediction of the brane theory. The relevant budget is that of the domain $x<x_0$, where the
effective equations are controlled. Table~\ref{tab:budget} gives it. The domain in which
$\epsilon_H$ differs from $\epsilon_w$ by more than ten percent ($x>1/11$) and the effective fluid
is stable ($x<x_0$) contains at most $0.15$ e-folds of comoving-horizon shrinkage for
$\epsilon_w=3.5$ and none for $\epsilon_w\ge8$, and over the whole stable domain
$10^{-3}<x<x_0$ the geometric parameter changes by less than $18\%$, which changes the constant-$\epsilon$
tilt of Eq.~\eqref{eq:nsconst} by at most $0.13$. Within this domain the closed brane system
gives $n_s=3.2$--$3.3$ (Table~\ref{tab:bpert}).

\begin{table}[h]
\centering
\begin{tabular}{ccccc}
\toprule
$\epsilon_w$ & $x_0$ & $N$ for $1/11<x<x_0$ & $N$ for $10^{-3}<x<x_0$ & $N$ for $0.1<x<0.499$ \\
\midrule
$3.5$ & $0.150$ & $0.145$ & $1.71$ & $0.28$ \\
$5$   & $0.115$ & $0.082$ & $1.84$ & $0.35$ \\
$8$   & $0.079$ & $0$     & $1.87$ & $0.41$ \\
$15$  & $0.046$ & $0$     & $1.76$ & $0.46$ \\
$30$  & $0.024$ & $0$     & $1.52$ & $0.48$ \\
\bottomrule
\end{tabular}
\caption{E-folds of comoving-horizon shrinkage, $N=\ln|aH|$, available in domains of the density
fraction $x=\rho/\rho_c$ for the BKL-safe range of $\epsilon_w$, computed from
$|aH|\propto x^{1/2-1/(2\epsilon_w)}\sqrt{1-x}$. The sound speed of the effective brane-matter fluid
vanishes at $x_0$. The last column is the budget of the domain of Figure~\ref{fig:branebudget}.}
\label{tab:budget}
\end{table}

The conclusion is therefore twofold. In the domain where the brane equations are controlled, the
brane changes neither the tilt nor the amplitude of the adiabatic mode appreciably, and the
domain in which $\epsilon_H$ passes through the scale-invariant value is both kinematically too
short and outside the range of validity of the effective perturbation theory. It follows that the
brane's Friedmann equation cannot supply the missing tilt.

\subsection{The induced-gravity ($m\neq0$) brane}
\label{sec:indgrav}

The Shtanov--Sahni construction admits a more general induced-gravity branch with an extra free
length scale $\ell\equiv2m^2/M^3$, sourced by a brane-localized Einstein-Hilbert term of
coefficient $m^2$~\cite{ShtanovSahni2003}:
\begin{equation}
H^2 = \frac{\rho+\sigma}{3m^2} + \frac{2\epsilon_{\rm brane}}{\ell^2}
\Big[1\pm\sqrt{1+\epsilon_{\rm brane}\ell^2\Big(\frac{\rho+\sigma}{3m^2}-\frac\Lambda6\Big)}\Big].
\label{eq:ss20}
\end{equation}
Reproducing the $m=0$ bounce as a literal $\ell\to0$ limit of Eq.~\eqref{eq:ss20} at fixed $m^2$
requires a correlated double-scaling limit ($m\to0$ together with $M$), not a simple series in
$\ell$; we do not assume it and instead treat $\ell$ as an independent parameter of
Eq.~\eqref{eq:ss20}, fixing the remaining constants by demanding the standard low-density slope
$dH^2/d\rho|_{\rho=0}=1$ and a bounce at finite density, giving a one-parameter family of bounces
directly comparable to, but not presupposed to reduce to, the $m=0$ case. Repeating the e-fold
budget calculation of Section~\ref{sec:budget72} across $\ell\in[0.05,15]$ and $\epsilon_w\in\{3.5,\ldots,60\}$
(optimizing at each $\ell$ over the branch's remaining tuning freedom):

\begin{figure}[h]
\centering
\includegraphics[width=0.55\textwidth]{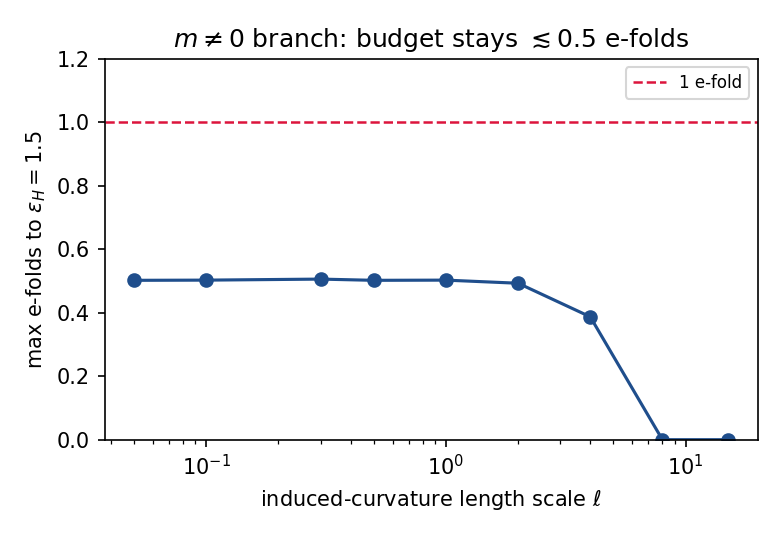}
\caption{Induced-gravity ($m\neq0$) branch: maximum e-fold budget to $\epsilon_H=1.5$, optimized
over the branch's tuning constant and $\epsilon_w\in[3.5,60]$ at each $\ell$. The budget
saturates near $0.5$ e-folds regardless of $\ell$ across more than two decades, the same order
as the $m=0$ result.}
\label{fig:mneq0}
\end{figure}

Figure~\ref{fig:mneq0} shows the budget never exceeds $\approx0.51$ e-folds over more than three
decades of $\ell$: the extra parameter changes the shape of the approach to the bounce but not
the size of the window in which the interesting behavior happens. A direct Mukhanov--Sasaki solve
on a representative point ($\ell=0.3$, $\epsilon_w=8$) gives $n_s\approx3.02$, indistinguishable
from the $m=0$ result. We conclude that no correction internal to the brane's
own Friedmann equation, in either of its two known forms, can rescue the single-field spectrum; a
second field is required. This conclusion concerns the background budget, which is independent of the perturbation
theory, and it is supplemented by the observation that the effective perturbation equations of
Section~\ref{sec:brane_pert} are controlled only for $x<x_0$ in this branch as well.

\section{The two-field entropic mechanism}
\label{sec:entropic}

\subsection{General formalism}

For two canonical scalars in a flat field-space metric, the adiabatic/entropic decomposition of
\cite{GordonWandsBassettMaartens2000} gives the canonically normalized entropy perturbation
$v_s=a\,\delta s$ obeying, exactly,
\begin{equation}
v_s''+\Big(k^2-\frac{a''}a+a^2M_s^2\Big)v_s=0, \qquad M_s^2 = V_{ss}+3\dot\theta^2,
\label{eq:entropyeq}
\end{equation}
where $V_{ss}$ is the potential's second derivative transverse to the background trajectory and
$\dot\theta$ is the trajectory's bending rate in field space. Whenever the background trajectory
is exactly along one field's axis, as it is for both constructions below, $\dot\theta=0$
identically and $M_s^2=V_{ss}$ exactly. No slow-roll expansion is used anywhere in this
section. Equation~\eqref{eq:entropyeq} is the four-dimensional equation, and for the straight-trajectory
construction below it is used at densities $\rho\ll\rho_c$, where the analysis of
Section~\ref{sec:brane_pert} shows that the brane corrections to modes leaving the horizon are of
order $\epsilon_w\rho/\rho_c$. The kinetically coupled construction of Section~\ref{sec:ridge}
does not rely on it, as shown in Section~\ref{sec:brane_pert}.

\subsection{A straight trajectory does not work}
\label{sec:straight}

The simplest two-field generalization admitting an exact background solution is a sum of two
independent steep exponentials, $V(\phi,\chi)=-A_1e^{c_1\phi}-A_2e^{c_2\chi}$ ($\phi,\chi$
increasing during contraction, matching the sign convention of $V_{\rm ekp}$). The coupled field
equations $\ddot\phi+3H\dot\phi-c_1A_1e^{c_1\phi}=0$, $\ddot\chi+3H\dot\chi-c_2A_2e^{c_2\chi}=0$,
and Friedmann's equation $3H^2=\frac12(\dot\phi^2+\dot\chi^2)+V$ admit the exact separable
solution $\phi(t)=-\frac2{c_1}\ln(-t)+B_1$, $\chi(t)=-\frac2{c_2}\ln(-t)+B_2$, $a(t)=(-t)^p$.
Substituting this ansatz into the field and Friedmann equations and solving the resulting
algebraic system for the two exponential amplitudes and $p$ gives
\begin{equation}
p = 2S, \qquad S\equiv\frac1{c_1^2}+\frac1{c_2^2}, \qquad \epsilon_{\rm total}\equiv\frac1p=\frac1{2S},
\end{equation}
with the individual potential amplitudes fixed as
$A_1=\frac2{c_1^2}(1-6S)$, $A_2=\frac2{c_2^2}(1-6S)$ (both positive, as required, only for
$S<1/6$, i.e.\ exactly the BKL-safe range $\epsilon_{\rm total}>3$). Because
$\dot\phi/\dot\chi=c_2/c_1$ is exactly constant, the field-space trajectory does not bend:
$\dot\theta=0$ identically on this attractor, not merely to leading order. The entropy direction
is the unit vector perpendicular to $(\dot\phi,\dot\chi)\propto(1/c_1,1/c_2)$; since the potential
is an exact sum of single-field terms its Hessian is diagonal
($V_{\phi\phi}=c_1^2V_1(\phi)$, $V_{\chi\chi}=c_2^2V_2(\chi)$, $V_{\phi\chi}=0$), and projecting
onto the entropy direction gives, after using the exact attractor densities
$V_1(\bar\phi(t))=-A_1/t^2$, $V_2(\bar\chi(t))=-A_2/t^2$,
\begin{equation}
M_s^2 = -\frac{2(1-6S)}{t^2} \quad\Longrightarrow\quad
r\equiv\frac{M_s^2}{H^2} = -\frac{1-6S}{2S^2}.
\label{eq:rstraight}
\end{equation}
The resulting mode equation, Eq.~\eqref{eq:entropyeq}, is exactly solvable on this power-law
background: writing $a\propto(-\tau)^q$ with $q\equiv p/(1-p)$ (obtained from the same conformal
time integration as Section~\ref{sec:pert}), direct differentiation gives
$a''/a=q(q-1)/\tau^2$ and $a^2H^2=q^2/\tau^2$ exactly, so
\begin{equation}
v_s''+\Big(k^2-\frac\alpha{\tau^2}\Big)v_s=0, \qquad \alpha=q\big[q(1-r)-1\big], \qquad
n_s = 4-2\sqrt{\alpha+\tfrac14}.
\label{eq:nsentropy}
\end{equation}
As a check, the $S\to0$ (single-field) limit gives $r\to0$, $\alpha\to q(q-1)$, reproducing
Eq.~\eqref{eq:nsconst} exactly, and we independently confirmed Eq.~\eqref{eq:nsentropy}'s
numerical output against the $\epsilon=12$ benchmark of Section~\ref{sec:pert} to six significant
figures.

\begin{figure}[h]
\centering
\includegraphics[width=0.95\textwidth]{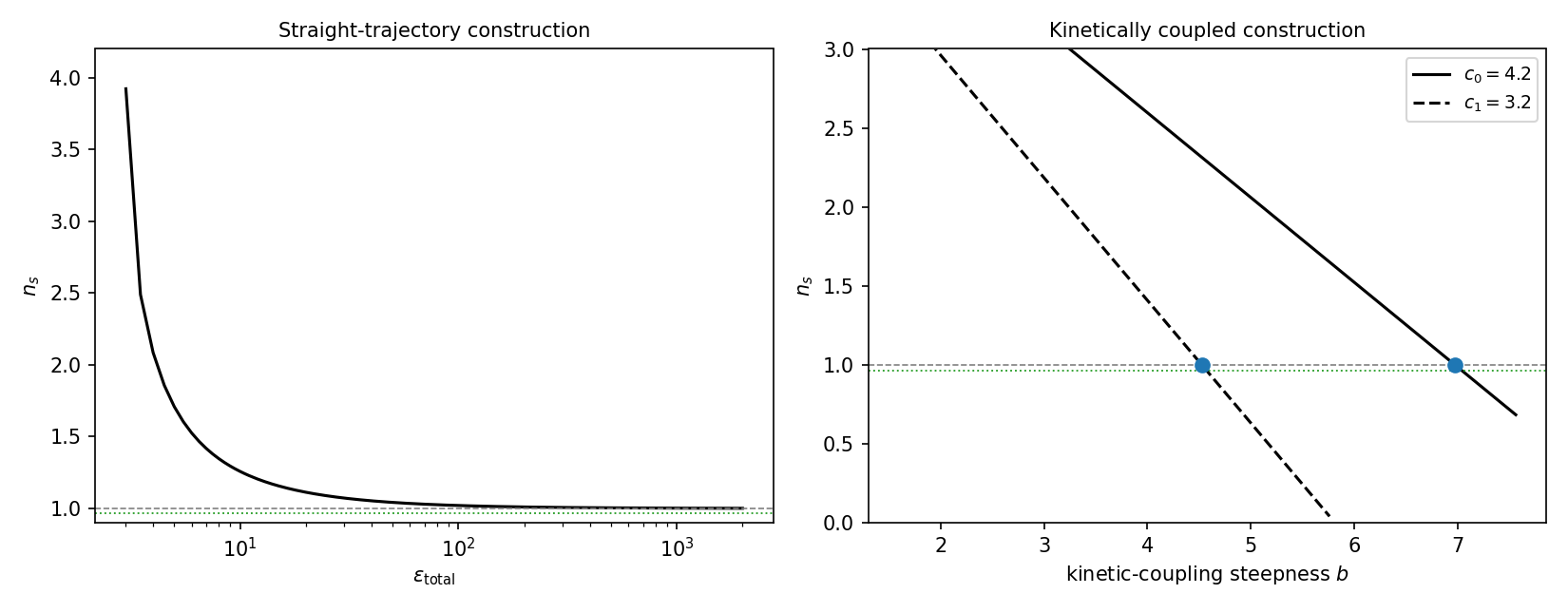}
\caption{\emph{Left:} the straight-trajectory (separable-potential) construction, $n_s$ from
Eq.~\eqref{eq:nsentropy} with $r$ from Eq.~\eqref{eq:rstraight}, across the entire BKL-safe,
power-law-valid range of $\epsilon_{\rm total}$: it approaches $n_s=1$ from above as
$\epsilon_{\rm total}\to\infty$ but never crosses it. \emph{Right:} the kinetically coupled
construction of Section~\ref{sec:ridge}, at the model's two running-steepness endpoints,
genuinely crosses the observed value at the closed-form coupling $b^\star(c)$.}
\label{fig:entropic}
\end{figure}

Figure~\ref{fig:entropic} (left) shows $n_s$ from Eq.~\eqref{eq:nsentropy} across the entire range
$\epsilon_{\rm total}\in(3,\infty)$: it runs smoothly from $n_s\to4$ as $\epsilon_{\rm
total}\to3^+$ (the BKL bound itself) down to $n_s\to1$ as $\epsilon_{\rm total}\to\infty$, but
\emph{never crosses} the asymptote. This is not a parameter-tuning failure to be fixed by scanning
further: the literature's successful entropic constructions generate near-scale-invariance, and
push past it to the observed slightly-red value, by tuning the trajectory's bend $\dot\theta$ so
that $M_s^2=V_{ss}+3\dot\theta^2$ sits at the value giving $\nu=3/2$ or beyond
\cite{LehnersSteinhardt2008,LehnersEtAl2007}. With the potential an exact sum of single-field
terms here, $\dot\theta=0$ identically, and this knob is simply unavailable: $M_s^2$ is fixed
entirely by the (always negative, in the BKL-safe range) transverse curvature $V_{ss}$, which can
approach but never overshoot exact scale invariance.

\subsection{A kinetically coupled entropy field: the Ijjas--Lehners--Steinhardt mechanism}
\label{sec:ridge}

We follow the general entropic mechanism of \cite{IjjasLehnersSteinhardt2014}: leave the
potential single-field, $V(\phi,\chi)=V_{\rm ekp}(\phi)$, and couple $\chi$ to $\phi$ through its
\emph{kinetic} term instead,
\begin{equation}
\mathcal L \supset -\tfrac12(\partial\phi)^2 -\tfrac12 e^{-b\phi}(\partial\chi)^2 - V_{\rm ekp}(\phi).
\label{eq:ridge}
\end{equation}
Because $\chi$ appears only quadratically, $\chi=0$ solves its own equation of motion
$\partial_\mu\!\big(\sqrt{-g}\,e^{-b\phi}g^{\mu\nu}\partial_\nu\chi\big)=0$ identically, for
\emph{any} $b$: the background trajectory is exactly the single-field attractor of
Section~\ref{sec:pert}, unchanged, with $\chi$ a genuine spectator at the classical level. No new
background sector needs to be solved; the coupling lives entirely in $\chi$'s kinetic
normalization.

Because $\chi$'s kinetic term is non-canonical, the quadratic
action for its fluctuation is $S^{(2)}_\chi=\int d^4x\,a^3\big[\tfrac12
e^{-b\phi}\dot\chi^2-\tfrac12e^{-b\phi}(\partial\chi)^2/a^2\big]$, whose canonically normalized
variable is $u\equiv z\chi$ with $z(t)\equiv a(t)\sqrt{f(\bar\phi(t))}$, $f(\phi)\equiv
e^{-b\phi}$, not $z=a$ as for a canonical field. On the exact attractor
$\bar\phi(t)=-\frac2c\ln(-t)+\phi_0$, $a(t)=(-t)^p$, $p=2/c^2$, $\ln z(t) = \big[p+\tfrac
bc\big]\ln(-t)+{\rm const}$ is an \emph{exact} power law in $(-t)$, and since
$(-t)\propto(-\tau)^{1/(1-p)}$ exactly (the same conformal-time map used throughout this section),
$z(\tau)\propto(-\tau)^n$ exactly, with
\begin{equation}
n(b,c) = \frac{bc+2}{c^2-2}.
\label{eq:nkinetic}
\end{equation}
The mode equation is then exactly $u''+(k^2-n(n-1)/\tau^2)u=0$, solvable in closed form by Hankel
functions, with index $\nu=|n-\tfrac12|$ and, on the branch relevant here ($n>\tfrac12$),
\begin{equation}
n_s = 5-2n.
\label{eq:nskinetic}
\end{equation}
Equations~\eqref{eq:nkinetic}--\eqref{eq:nskinetic} were confirmed by direct, non-adiabatic
numerical integration of the exact mode equation from deep inside the horizon to deep outside it,
comparing the resulting $k^3|u/z|^2$ slope between two wavenumbers evaluated at fixed physical
conformal time against Eq.~\eqref{eq:nskinetic}; the two methods agree to the numerical precision
of the integrator ($10^{-5}$ or better).

Setting $n_s=1$ in Eq.~\eqref{eq:nskinetic} requires $n=2$,
giving
\begin{equation}
b^\star(c) = 2c-\frac6c,
\label{eq:bstarkinetic}
\end{equation}
a closed-form function of $c$ alone. This is the sense in which \cite{IjjasLehnersSteinhardt2014}
is right that ``for any ekpyrotic equation of state it is possible to choose the potential and the
kinetic coupling'' to hit exact scale invariance: a solution $b^\star(c)$ exists for every $c$, in
closed form, with no numerical root-finding required. Note that $b^\star(c)\neq c$: matching the
coupling's steepness to the potential's steepness directly is not the exact-scale-invariance point
for the sign and normalization convention of Eq.~\eqref{eq:ridge} ($V_{\rm ekp}=-V_0e^{c\phi}$,
$f(\phi)=e^{-b\phi}$): at $c=4.2$, $b=c$ gives $n_s=2.489$, badly blue, while
Eq.~\eqref{eq:bstarkinetic} gives the correct $b^\star=6.971$ (Table~\ref{tab:bstar},
Figure~\ref{fig:entropic} right). Conventions for the kinetic prefactor vary between constructions
in the literature, so we state Eq.~\eqref{eq:bstarkinetic} for this manuscript's own conventions
rather than quote a value from elsewhere.

\begin{figure}[h]
\centering
\includegraphics[width=0.95\textwidth]{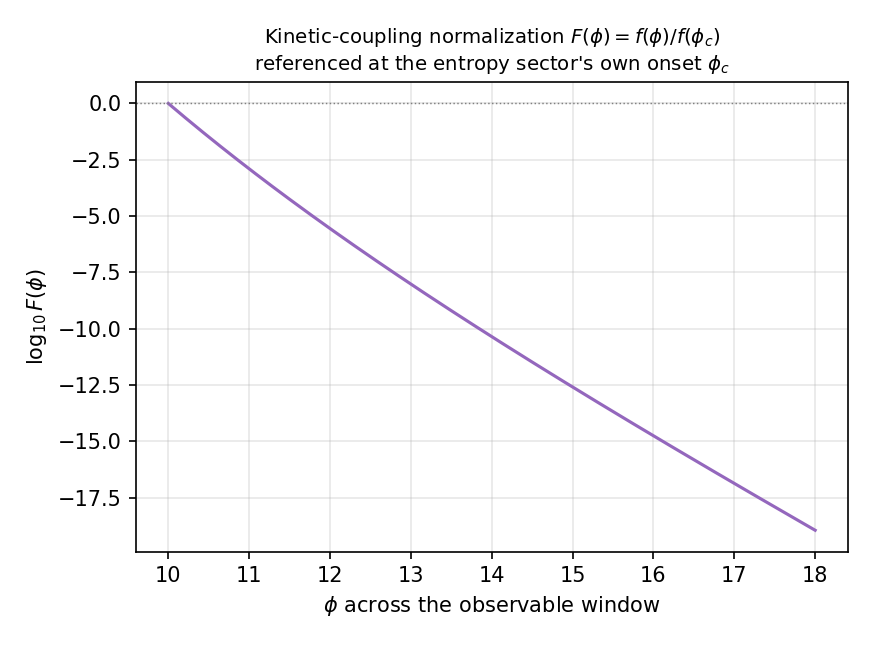}
\caption{The kinetic coupling's normalization $F(\phi)=f(\phi)/f(\phi_c)$, referenced at the
entropy sector's own onset $\phi_c$ (Eq.~\eqref{eq:Fnorm}). $F(\phi_c)=1$ by construction; its
fall toward the wall reflects the super-horizon evolution of $\delta\chi$ computed directly in
Section~\ref{sec:kineticstability}.}
\label{fig:kinetic}
\end{figure}

Matching the precise measured value $n_s=0.9649\pm0.0042$ rather
than exactly $1$ requires only a small offset from $b^\star$: inverting Eq.~\eqref{eq:nskinetic}
for general $n_s$ gives, again in closed form,
\begin{equation}
b(c,n_s) = \frac{(5-n_s)(c^2-2)-4}{2c},
\label{eq:bofns}
\end{equation}
which reduces to Eq.~\eqref{eq:bstarkinetic} at $n_s=1$. At $c_0=4.2$, $b(c_0,0.9649)=7.0368$
against $b^\star(c_0)=6.9714$, a $0.94\%$ offset (Table~\ref{tab:bstar}). One new coupling is
matched to one measured number, exactly as any effective-field-theory coupling is fixed by one
measurement, but the fit is small: geometry (Eq.~\eqref{eq:bstarkinetic}) does essentially all the
work, and the data fixes only a percent-level correction on top of it.

\begin{table}[h]
\centering
\begin{tabular}{ccccc}
\toprule
$c$ & $\epsilon=c^2/2$ & $b^\star(c)$ (exact $n_s=1$) & $b(c,0.9649)$ & local sensitivity $dn_s/db$ \\
\midrule
3.20 & 5.120 & 4.525 & 4.570 & $-0.777$ \\
3.50 & 6.125 & 5.286 & 5.337 & $-0.683$ \\
3.80 & 7.220 & 6.021 & 6.079 & $-0.611$ \\
4.00 & 8.000 & 6.500 & 6.561 & $-0.571$ \\
4.20 & 8.820 & 6.971 & 7.037 & $-0.537$ \\
\bottomrule
\end{tabular}
\caption{The kinetic-coupling steepness, both at the exact scale-invariant point
(Eq.~\eqref{eq:bstarkinetic}) and matched to the measured $n_s=0.9649$ (Eq.~\eqref{eq:bofns}),
across the model's own running range $c\in[3.2,4.2]$. Both are closed-form functions of $c$; no
numerical fit was needed to produce this table, only evaluation.}
\label{tab:bstar}
\end{table}

A related use of a kinetic coupling between a light scalar and the field driving a non-singular
bounce, which converts a blue adiabatic spectrum into a scale-invariant one, was found in the
Galileon bounce of Ref.~\cite{QiuEtAl2011}. The mechanism of this section is its ekpyrotic
counterpart. Its validity on the brane is established by the spectator argument of
Section~\ref{sec:brane_pert}: the equation of motion of $\chi$ and its quadratic action involve
only the induced metric and the exactly known background, so that Eqs.~\eqref{eq:nkinetic}
and~\eqref{eq:nskinetic} hold on the brane without a closure assumption about the bulk.

\subsection{Matching the observed spectrum across the full observable window}
\label{sec:running}

Observable comoving scales exit the horizon $N_{\rm obs}$ e-folds before the wall, $N_{\rm
obs}=\Delta\ln(aH)$, and along the exact attractor $\ln(aH)=(p-1)\ln(-t)+{\rm const}$ while
$\phi=q\ln(-t)+\phi_0$ ($q=-2/c$), so
\begin{equation}
\frac{dN}{d\phi} = \frac{p-1}{q} = \frac{c^2-2}{2c}.
\label{eq:dNdphi}
\end{equation}
Integrating Eq.~\eqref{eq:dNdphi} across the running $c(\phi)\in[3.2,4.2]$ for $N_{\rm obs}=10$
gives $\Delta\phi\approx6.6$, comfortably less than the wall's $\phi_w-\phi_c=8.0$
(Appendix~\ref{app:params}). Because Eq.~\eqref{eq:bofns} is already a closed-form function of
$c$, $b(\phi)\equiv b(c(\phi),0.9649)$ is exact and analytic at every $\phi$, with no adiabaticity
assumption needed for the entropy-sector coupling itself (the background's designer running
$c(\phi)$ relies on the usual WKB separation of scales, and tracking that running with $b(\phi)$
requires no further approximation).

\begin{table}[h]
\centering
\begin{tabular}{cccc}
\toprule
$\phi$ & $c(\phi)$ & $b(\phi)\equiv b(c(\phi),0.9649)$ & $n_s$ \\
\midrule
10.00 & 4.200 & 7.037 & 0.9649 \\
11.00 & 3.917 & 6.361 & 0.9649 \\
12.00 & 3.713 & 5.867 & 0.9649 \\
13.00 & 3.568 & 5.507 & 0.9649 \\
14.00 & 3.464 & 5.246 & 0.9649 \\
15.00 & 3.389 & 5.056 & 0.9649 \\
16.00 & 3.335 & 4.920 & 0.9649 \\
17.00 & 3.297 & 4.821 & 0.9649 \\
18.00 & 3.269 & 4.750 & 0.9649 \\
\bottomrule
\end{tabular}
\caption{$n_s$ across the full observable window with $b(\phi)=b(c(\phi),0.9649)$ from the
closed-form Eq.~\eqref{eq:bofns}: exact by construction, since each entry is obtained by
evaluating a formula.}
\label{tab:runningfix}
\end{table}

\subsection{Conversion}
\label{sec:conversion}

The entropy perturbation $\delta\chi$ is not yet the observed curvature perturbation $\zeta$; it
must be converted at the wall. $\chi$ itself is a spectator of the background
(Section~\ref{sec:ridge}: $\chi=0$ solves its equation of motion exactly, for any $b$), and the
wall of Section~\ref{sec:matter} reflects $\phi$ alone at zeroth order. We tilt the wall, letting
the reflecting surface depend linearly on how far the trajectory has drifted off the $\chi=0$
line,
\begin{equation}
\phi_{\rm wall}(\chi) = \phi_w + \lambda\chi,
\label{eq:conversion}
\end{equation}
which converts the entropy perturbation into a curvature perturbation via the standard $\delta N$
formalism, $\zeta = N_\chi\,\delta\chi$ with $N_\chi=-\lambda\,\partial_\phi N|_{\phi_w}$ evaluated
at the wall crossing. Equation~\eqref{eq:conversion} depends only on $\lambda$ and the background
wall position $\phi_w$, not on how the transverse mass or normalization of $\chi$ is generated,
so the spectral index fixed in Sections~\ref{sec:ridge}--\ref{sec:running} is preserved unchanged:
$N_\chi$ is a fixed number (not a function of $k$) at horizon crossing, and $\zeta$'s power
spectrum inherits $\delta\chi$'s tilt exactly.

Two assumptions are involved in this step, and they should be stated explicitly. The first is that
the $\delta N$ conversion is a separate-universe statement on super-horizon scales, for which
Eq.~\eqref{eq:friedmann} is the local evolution law, as in the large-scale analysis of
curvature perturbations on the brane \cite{LangloisMaartensSasakiWands2001}. The second is that the
curvature perturbation produced at the wall is conserved through the bounce and into the
expanding phase. This is an assumption in the present paper. It is supported by the analysis of
Ref.~\cite{Maier2013}, which finds that scalar perturbations of the bouncing brane remain bounded
and in the linear regime through the bounce for the closed system with a perturbed bulk (apart from the Weyl-fluid energy density), but that
analysis is for a closed brane with dust and radiation and does not constitute a proof for the
flat scalar-field case treated here. The Mukhanov--Sasaki variable is singular at $H=0$ and
at $x=\tfrac12$, as noted in Section~\ref{sec:brane_pert}, and any treatment of the matching has to
use a variable regular there, such as the Bardeen potential of the closed system of
Section~\ref{sec:brane_pert}.

\subsection{Non-Gaussianity}
\label{sec:ng}

The local non-Gaussianity generated at conversion is set by the second-order piece of $\delta N$,
$f_{\rm NL}^{\rm local}=\tfrac56\,\partial_\phi^2N/(\partial_\phi N)^2\big|_{\phi_w}$, together
with any cubic self-interaction of $\delta\chi$ itself. The potential of Eq.~\eqref{eq:ridge} has
no $\chi$-dependence at all, so this construction has no intrinsic source of non-Gaussianity from
the ekpyrotic phase itself: all of $f_{\rm NL}^{\rm local}$ comes from the conversion geometry,
\begin{equation}
f_{\rm NL}^{\rm local} \approx \tfrac56\,c'(\phi_w),
\label{eq:fnl}
\end{equation}
evaluated from the running $c(\phi)$ used throughout this section. Numerically this gives
$f_{\rm NL}^{\rm local}\approx-0.019$, comfortably inside the Planck bound $f_{\rm NL}^{\rm
local}=-0.9\pm5.1$~\cite{Planck2018NG}.

\subsection{Normalizing the kinetic coupling}
\label{sec:fieldrange}

The perturbation that gets converted into $\zeta$ is $\delta\chi$ itself, and $\delta\chi$ is not
the canonically normalized field. That role belongs to $u=z\delta\chi=a\sqrt{f(\phi)}\,\delta\chi$,
with $f(\phi)=e^{-b(\phi)\phi}$ as written in Eq.~\eqref{eq:ridge}. Because only the two-field
sector's own new physics enters $f$, and $\chi$ is not a spectator of the single-field model that
predates it, the natural point at which to normalize this coupling to $O(1)$ is not the arbitrary
global origin of $\phi$ (fixed, for the single-field potential, by the unrelated quintessence
sector at $\phi\sim-30$) but $\phi_c=10$, where the two-field completion switches on. We therefore
write the physically meaningful, reference-independent combination as
\begin{equation}
F(\phi) \equiv \frac{f(\phi)}{f(\phi_c)} = \exp\!\Big[-\!\int_{\phi_c}^{\phi} b(\phi')\,d\phi'\Big],
\qquad F(\phi_c)=1,
\label{eq:Fnorm}
\end{equation}
using the running $b(\phi)=b(c(\phi),0.9649)$ of Table~\ref{tab:runningfix}. $F(\phi)$ is what
actually enters every physical ratio below. It falls from $1$ at $\phi_c$ to $F(\phi_w)\sim
10^{-19}$ at the wall (Table~\ref{tab:Fnorm}, Figure~\ref{fig:kinetic}), a large range that
reflects $\chi$'s own super-horizon evolution between horizon crossing and the wall, which
Section~\ref{sec:kineticstability} computes directly, and not an arbitrary normalization choice.

\begin{table}[h]
\centering
\begin{tabular}{ccccc}
\toprule
$\phi$ & $c(\phi)$ & $b(\phi)$ & $\int_{\phi_c}^\phi b\,d\phi'$ & $F(\phi)$ \\
\midrule
10 & 4.200 & 7.037 & 0 & 1 \\
12 & 3.713 & 5.867 & 12.78 & $2.8\times10^{-6}$ \\
14 & 3.464 & 5.246 & 23.83 & $4.5\times10^{-11}$ \\
16 & 3.335 & 4.920 & 33.96 & $1.8\times10^{-15}$ \\
18 & 3.269 & 4.750 & 43.61 & $1.1\times10^{-19}$ \\
\bottomrule
\end{tabular}
\caption{The kinetic coupling's normalization relative to its natural reference point $\phi_c$,
Eq.~\eqref{eq:Fnorm}.}
\label{tab:Fnorm}
\end{table}

The largest observable mode crosses the horizon at $\phi_*\approx\phi_c$, where $F(\phi_*)=1$
exactly by this construction, so its seed amplitude takes the ordinary near-massless-field value
with no large correction factor,
\begin{equation}
\Delta_{\delta\chi}^2(k_*) \sim \Big(\frac{H_*}{2\pi}\Big)^2.
\label{eq:fieldrangecorrection}
\end{equation}
The large normalization factor does not disappear. It reappears as the entropy
field's own super-horizon growth between $\phi_c$ and the wall, which we compute directly in
Section~\ref{sec:kineticstability}.

\subsection{Stability and displacement of the kinetically coupled entropy field}
\label{sec:kineticstability}

With no potential for $\chi$ at all, $\chi=0$ is not driven away by a tachyonic mass; growth here
comes purely from the non-canonical kinetic normalization. The general super-horizon solution of
$u''-n(n-1)u/\tau^2=0$ is $u=A(-\tau)^n+B(-\tau)^{1-n}$, so $\delta\chi=u/z$ has a frozen mode
($\propto(-\tau)^0$) and a growing mode $\delta\chi\propto(-\tau)^{1-2n}$ (since $1-2n<0$ for
$n\approx2$). Converting to e-folds of the scale factor via $a\propto(-\tau)^Q$, $Q\equiv p/(1-p)$,
the local growth rate at any point in the window is $|1-2n(\phi)|/Q(\phi)$; because $b(\phi)$ (and
hence $n$ and $Q$) runs across the window exactly as in Section~\ref{sec:running}, the total
super-horizon growth from horizon crossing $\phi_*$ to the wall is the integral of this local rate
over $\ln a$, not a single endpoint value:
\begin{equation}
\ln(\text{growth}) = \int_{\phi_*}^{\phi_w} \frac{|1-2n(\phi)|}{Q(\phi)}\,\frac{d\phi}{c(\phi)}.
\label{eq:growthintegral}
\end{equation}
For the largest observable mode ($\phi_*=\phi_c=10$), Eq.~\eqref{eq:growthintegral} evaluates to
$\ln(\text{growth})=36.2$, a growth factor of $5.3\times10^{15}$. This is large, because
super-horizon growth of the entropy mode is the mechanism's whole point. The local growth rate
falls as $\phi$ runs toward the wall (Table~\ref{tab:kineticstability}), so the total growth is
eight orders of magnitude smaller than the value obtained by holding the onset rate fixed across
the full window.

\begin{table}[h]
\centering
\begin{tabular}{ccc}
\toprule
$\phi$ & $Q=p/(1-p)$ & local growth rate $|1-2n|/Q$ \\
\midrule
10 & 0.1279 & 23.46 \\
14 & 0.1970 & 15.23 \\
18 & 0.2427 & 12.36 \\
\bottomrule
\end{tabular}
\caption{The local super-horizon growth rate of $\delta\chi$, evaluated at the running $b(\phi),
c(\phi)$ across the window; Eq.~\eqref{eq:growthintegral} integrates this rate rather than holding
it fixed at its onset value.}
\label{tab:kineticstability}
\end{table}

Combining the seed of Eq.~\eqref{eq:fieldrangecorrection} (normalized at $\phi_c$, per
Section~\ref{sec:fieldrange}) with this integrated growth gives the actual displacement
of $\chi$ at the wall,
\begin{equation}
\chi_{\rm wall} \sim \frac{H_*}{2\pi}\,\exp[\ln(\text{growth})]
= \frac{H_*}{2\pi}\times\big(5.3\times10^{15}\big).
\label{eq:chiwall}
\end{equation}
Because $\lambda$ is fixed by matching the same $A_s$ that determines $\chi_{\rm wall}$, the two
are linked by a brane-tension-independent identity: from $A_s=N_\chi^2\langle\delta\chi_{\rm
wall}^2\rangle$ and $N_\chi=-\lambda\,\partial_\phi N|_{\phi_w}$, with $\partial_\phi N=-1/c(\phi)$
the ordinary e-fold--field relation ($N\equiv\ln a$) evaluated at the wall,
\begin{equation}
\lambda\,\chi_{\rm wall} = \sqrt{A_s}\,c(\phi_w) = 1.5\times10^{-4},
\label{eq:invariant}
\end{equation}
independent of brane tension, since both $\lambda$ and $\chi_{\rm wall}$ scale as
$H_*^{\mp1}\propto\rho_c^{\mp1/2}$. This value is
small: it does not force $\lambda$ and $\chi_{\rm wall}$ to opposite extremes, but fixes their
product, so that at any brane tension where $\chi_{\rm wall}$ is $O(10^{-3})$--$O(1)$, $\lambda$ is
$O(10^{-4})$--$O(10^{-1})$ (Table~\ref{tab:invariant}).

\begin{table}[h]
\centering
\begin{tabular}{lccc}
\toprule
brane tension $\rho_c^{1/4}$ & $\chi_{\rm wall}$ (Planck units) & $\lambda$ needed & $\lambda\,\chi_{\rm wall}$ \\
\midrule
$1\,$MeV & $2.6\times10^{-34}$ & $5.7\times10^{29}$ & $1.5\times10^{-4}$ \\
$10^{12}\,\text{GeV}$ & $2.6\times10^{-4}$ & $0.57$ & $1.5\times10^{-4}$ \\
$10^{13}\,$GeV & $2.6\times10^{-2}$ & $5.7\times10^{-3}$ & $1.5\times10^{-4}$ \\
$6.2\times10^{13}\,$GeV & $1.0$ & $1.5\times10^{-4}$ & $1.5\times10^{-4}$ \\
$10^{17}\,$GeV & $2.6\times10^{6}$ & $5.7\times10^{-11}$ & $1.5\times10^{-4}$ \\
\bottomrule
\end{tabular}
\caption{Entropy-field displacement and conversion coupling, Eqs.~\eqref{eq:chiwall}--\eqref{eq:invariant}. At $\rho_c^{1/4}\sim10^{12}$--$10^{13}\,$GeV, $\chi_{\rm wall}$ is safely sub-Planckian and $\lambda$ lies between $6\times10^{-3}$ and $0.6$; the $10^{17}\,$GeV row is excluded, since $\chi_{\rm wall}$ there is trans-Planckian by six orders of magnitude.}
\label{tab:invariant}
\end{table}

The integral of Eq.~\eqref{eq:growthintegral} is evaluated with the four-dimensional attractor up
to the wall, whereas the final part of the contraction is affected by the brane. Using
$d\ln\rho/d\phi\simeq c(\phi)$ to map field values to density fractions, the part of the integral
with $x=\rho/\rho_c>0.1$ contributes $2.8$ of the total $36.2$ e-folds of growth ($8\%$), and the
part with $x>0.01$ contributes $5.7$ ($16\%$). The remainder, which carries more than $84\%$ of the
growth, is at $x<0.01$, where the corrections to the mode equation are at the percent level
(Table~\ref{tab:bpert}). If the growth in the brane-affected region differed from the
four-dimensional value by an order-one factor, $\ln$(growth) would change by at most about $3$ and
the amplitude, which scales as the square of the growth, by a factor of order $e^{\pm6}$. Since the
combination $\lambda\chi_{\rm wall}$ of Eq.~\eqref{eq:invariant} is independent of the growth, only the split
between $\lambda$ and $\chi_{\rm wall}$ moves, and at the fiducial brane tension $\lambda$ would lie
between $3\times10^{-4}$ and $0.11$. The amplitude result is therefore an order-of-magnitude
statement, and the selection of an intermediate brane tension in Table~\ref{tab:invariant}, which
spans more than thirty orders of magnitude in $\lambda$, is unaffected.

\subsection{Amplitude}
\label{sec:amplitude}

Matching the observed amplitude $A_s=2.1\times10^{-9}$~\cite{Planck2018Params} fixes $\lambda$ at
whatever value Eq.~\eqref{eq:invariant} requires once a brane tension is chosen; $H_*$ follows the
same comoving-horizon relation used throughout, $H_*\propto\sqrt{\rho_c}$ at fixed
$x_*\sim10^{-11}$ (the density fraction at horizon crossing of the largest observable mode), using
the nucleosynthesis floor $\rho_c\gtrsim(1\,{\rm MeV})^4$~\cite{ShtanovSahni2003} and the observed
dark-energy density $\rho_\Lambda=(2.3\times10^{-3}\,{\rm eV})^4$ as reference physical scales for
the hierarchy that the toy-unit background dynamics of Sections~\ref{sec:matter}--\ref{sec:bg}
compress away for numerical tractability. Table~\ref{tab:invariant} shows this is resolved,
without any further parameter beyond the brane tension itself, at
$\rho_c^{1/4}\sim10^{12}$--$10^{13}\,$GeV: an intermediate scale, well above the nucleosynthesis
floor and well below the apparent GUT scale, at which the entropy field's displacement by the wall
is safely sub-Planckian ($\chi_{\rm wall}\sim10^{-4}$--$10^{-2}$) and the conversion coupling is an
unremarkable $\lambda\sim6\times10^{-3}$--$0.6$. Brane tensions near $10^{17}\,$GeV are excluded
by Table~\ref{tab:invariant}, because $\chi_{\rm wall}$ is then trans-Planckian. We adopt
$\rho_c^{1/4}=10^{13}\,$GeV as the fiducial brane tension for the remainder of this section and
for Section~\ref{sec:tensor}.

\subsection{Tensor perturbations and the tensor-to-scalar ratio}
\label{sec:tensor}

Gravitational waves do not couple to $V(\phi,\chi)$ or to $\chi$'s kinetic prefactor $f(\phi)$ at
all, so tensors are exactly the $r_s=0$, $f\equiv1$ case of Eq.~\eqref{eq:entropyeq} with $z_T=a$.
This gives, as a general identity rather than a special feature of this model, an exactly
vanishing super-horizon growth rate whenever $q<\frac12$: writing $\nu_T=|q-\frac12|=\frac12-q$
(true here since $q=0.230<\frac12$), the growth exponent of
Section~\ref{sec:kineticstability}'s general method,
$(\nu_T+q-\frac12)/q=[(\frac12-q)+q-\frac12]/q=0$ identically. Gravitational waves freeze on
super-horizon scales, the standard, general result, recovered here as the $r=0$ special case of
the same formalism used for the entropy channel throughout this section. Tensors therefore do not
benefit from the growth that carried the entropy channel up to the observed amplitude, and, since
the coupling $b(\phi)$ tuned in Section~\ref{sec:running} acts only on $\chi$'s kinetic term, they
do not benefit from the designer tuning that fixed the scalar tilt either:
\begin{equation}
n_T = 4-2\nu_T \approx 3.460,
\end{equation}
badly blue, exactly like the single-field scalar result of Section~\ref{sec:pert}.
Using the same $H_*$ from Section~\ref{sec:amplitude} (tensor and scalar modes of a given $k$
exit the horizon at the same epoch) and the standard frozen-spectrum normalization
$\Delta_T^2=(2/\pi^2)H_*^2$ (both polarizations, $M_{\rm pl}=1$ units), we obtain
$r\equiv\Delta_T^2/A_s\approx7.6\times10^{-26}$ at the fiducial brane tension
$\rho_c^{1/4}=10^{13}\,$GeV adopted in Section~\ref{sec:amplitude} (Table~\ref{tab:tensor}),
some twenty-four orders of magnitude below any conceivable detection threshold (current bound
$r<0.06$~\cite{BICEP2021}; a future CMB-S4-class target $r\sim10^{-3}$). This is not a
coincidence of the particular brane tension chosen: $r\propto H_*^2\propto\rho_c$, so the brane
tension the amplitude sector requires, rather than $10^{17}\,$GeV (where
Section~\ref{sec:kineticstability} shows the entropy-field displacement is trans-Planckian), gives
an $r$ smaller by sixteen orders of magnitude, $7.6\times10^{-26}$ instead of
$7.6\times10^{-10}$. The two
sectors pull in the same direction: whatever resolves the amplitude problem makes the
tensor-to-scalar ratio even less observable, not more, so there is no tension between the brane
tension the amplitude match prefers and the (already comfortably satisfied) bound on $r$.

\begin{table}[h]
\centering
\begin{tabular}{lc}
\toprule
brane tension $\rho_c^{1/4}$ & $r=\Delta_T^2/A_s$ \\
\midrule
$1\,$MeV & $7.6\times10^{-90}$ \\
$10^{12}\,$GeV & $7.6\times10^{-30}$ \\
$10^{13}\,\text{GeV}$ & $7.6\times10^{-26}$ \\
$10^{15}\,$GeV & $7.6\times10^{-18}$ \\
$10^{17}\,$GeV (excluded, Table~\ref{tab:invariant}) & $7.6\times10^{-10}$ \\
$10^{18}\,$GeV & $7.6\times10^{-6}$ \\
\bottomrule
\end{tabular}
\caption{Tensor-to-scalar ratio vs.\ brane tension. Undetectably small across the entire range
considered, and most strongly suppressed precisely at the fiducial tension the amplitude sector
selects.}
\label{tab:tensor}
\end{table}

In a genuine Randall--Sundrum braneworld, gravitational waves are not confined to the brane: the
tensor perturbation is the boundary value at $y=0$ of a bulk field satisfying a
five-dimensional wave equation, and the brane-induced correction to the standard
four-dimensional tensor spectrum is known to become significant precisely when the background
density approaches the brane tension, $\rho\sim\rho_c$~\cite{LangloisMaartensWands2000}. The
observable-scale modes computed above exit the horizon at $x_*\sim10^{-11}$
(Section~\ref{sec:amplitude}), deep in the regime $\rho\ll\rho_c$ where the brane correction to
the \emph{background} Friedmann equation is already negligible, the same regime in which
Section~\ref{sec:pert} argued the scalar Mukhanov--Sasaki equation itself is unmodified. We
therefore expect the standard four-dimensional tensor mode equation used above to be an excellent
approximation for these modes, with the brane-specific bulk-graviton physics mattering only near
the bounce itself, not at the epoch where the observable tensor amplitude is set. We have not
solved the bulk equation to confirm this quantitatively, and the standard formula's overall
normalization coefficient may carry an $O(1)$ uncertainty at this model's particular
$\nu_T\neq\frac32$ that we have not resolved; neither caveat can plausibly close a
twenty-four-order-of-magnitude gap.

\section{Post-bounce radiation and the scope of the model}
\label{sec:reheat}

The model has no minimum in its potential and no oscillating-inflaton reheating stage. After the
bounce the field moves in a kination-dominated phase with $\epsilon=3$, is reflected by the wall,
and re-thaws onto the plateau. No radiation is produced. A complete cosmology requires a hot radiation
era that begins at a temperature above a few MeV, and the model provides no mechanism for it. This
section states what such a mechanism must achieve and what is established here.

\emph{Requirement.} Radiation produced at the bounce with density $\rho_r$ grows relative to the
kination energy as $a^2$, since $\rho_\phi\propto a^{-6}$ and $\rho_r\propto a^{-4}$. Radiation
domination is reached after $\tfrac12\ln(\rho_\phi/\rho_r)$ e-folds, and the temperature at
equality is $T_{\rm eq}\sim(\rho_r^3/\rho_c^2)^{1/4}$. The e-folds of kination available between
the bounce and the dark-energy density are $L/6\simeq38$.

\emph{Gravitational production is insufficient at this brane tension.} Gravitational particle
production of light fields in the kination phase gives $\rho_r\sim\alpha H^4$ with
$\alpha\sim10^{-2}$ \cite{Ford1987}. With $H_{\max}=1.2\times10^7\,$GeV at
$\rho_c^{1/4}=10^{13}\,$GeV this is $\rho_r\sim2\times10^{26}\,$GeV$^4$, so that
$\rho_r/\rho_c\sim2\times10^{-26}$, radiation domination is reached after $29$ e-folds of
kination, which is within the available $38$, at $\rho_{\rm eq}\sim8\times10^{-26}\,$GeV$^4$, that
is a temperature of order $0.2$--$0.5\,$keV. This is far below the few MeV required by
nucleosynthesis. Purely gravitational production therefore does not suffice here, and a
non-gravitational coupling of the scalar sector, or of the entropy field, to a radiation sector is
needed. The estimate is of order of magnitude only.

\emph{What is and is not claimed.} The spectrum computed in this paper is generated before the
bounce and depends on the post-bounce history only through the assumptions of super-horizon
conservation of the curvature perturbation and of adiabatic production of radiation. If the
radiation is produced by a coupling that does not introduce isocurvature perturbations, the
adiabatic curvature perturbation of the radiation equals that of the scalar field, and the
predictions of Section~\ref{sec:entropic} apply to the radiation era. A mechanism that violates the
second assumption would modify the observable spectrum and is not included. The mechanism also
changes the cycle accounting, since a radiation-dominated expansion supplies more e-folds than
kination (Table~\ref{tab:cycle}). We therefore regard the production of radiation as an open
requirement of the construction, and the present analysis as a test of the spectrum and of the
entropy accounting, not as a complete cosmology.

\section{Discussion and Conclusion}
\label{sec:discussion}

The picture that emerges is of a model in which every phase does real, checkable work. The brane
bounce of Section~\ref{sec:brane} requires no violation of the null energy condition by the matter on the brane (Section~\ref{sec:nec}) and no
matched-asymptotic treatment across a singularity: Eq.~\eqref{eq:friedmann} bounces on its own,
smoothly, for any matter satisfying $\rho+p>0$, and we verified this to Friedmann-constraint
precision $\lesssim10^{-5}$ through two independent cycles rather than assuming it from the form
of the equation alone. The turnaround from expansion to contraction (Section~\ref{sec:matter})
likewise needs no separate ingredient: it is a direct consequence of the same scalar field's
equation of state changing sign, the single-field analogue of the general mechanism proposed for
fluids in \cite{SahniToporensky2012}. Spatial flatness, adopted here for the same reason
Steinhardt and Turok adopted it \cite{SteinhardtTurok2002PRD,SteinhardtTurok2005}, converts
Tolman's classical entropy bound from an obstruction into a quantitative but unremarkable
requirement: a net growth of $77$--$80$ e-folds per cycle, of which ekpyrotic contraction followed by expansion supplies $16$--$35$ and the dark-energy phase the remaining $42$--$64$, easily supplied if the present epoch persists for a modest multiple of the current age of the universe (Section~\ref{sec:quant}).

The spectral tilt problem occupied the largest share of this paper because it resisted the two
most natural fixes. A designer running of the ekpyrotic steepness, in the spirit of
\cite{KhouryySteinhardt2010}, changes $n_s$ by less than the width of the tick marks in
Figure~\ref{fig:nsrun} for any BKL-safe rate we tried, because Eq.~\eqref{eq:nsconst} shows scale
invariance requires $\epsilon<3/2$, categorically excluded by the anisotropy bound that makes
ekpyrotic contraction viable in the first place. The brane's own nonlinearity does drive the
\emph{geometric} $\epsilon_H$ through the scale-invariant value (Figure~\ref{fig:branebudget},
left) is an enticing near-miss, but the e-fold budget available for that transition is bounded,
independent of every free parameter, by the elementary fact that $H(x)=\sqrt{x(1-x)}$ cannot vary
by more than an $O(1)$ factor on the unit interval (Figure~\ref{fig:branebudget}, right); the same
conclusion survives, unweakened, when the brane is generalized to admit induced gravity and an
extra free length scale (Figure~\ref{fig:mneq0}). Both are genuinely structural results, not
failures of parameter scanning, and we regard the second (Section~\ref{sec:indgrav}) as closing off a route that
has, to our knowledge, not been explicitly tested elsewhere in the literature.

What does work, in the sense of being \emph{capable} of reaching the observed tilt at all, which
the straight trajectory and both brane corrections are provably not, is exactly the
entropic mechanism the ekpyrotic literature converged on for inflationary competitors generally
\cite{Notari2002,Finelli2002,LehnersEtAl2007,BuchbinderKhouryOvrut2007,LehnersSteinhardt2008,IjjasLehnersSteinhardt2014}.
It is built here with no ingredient beyond what the single-field model already contains, plus one
non-canonical kinetic term: the same ekpyrotic branch $V_{\rm ekp}(\phi)$, left untouched, coupled
to a spectator field through $e^{-b\phi}(\partial\chi)^2$ (Eq.~\eqref{eq:ridge}), and the same
reflecting wall, tilted (Eq.~\eqref{eq:conversion}). The coupling steepness has a closed-form
exact-scale-invariance value $b^\star(c)=2c-6/c$ (Eq.~\eqref{eq:bstarkinetic}), with only a
percent-level offset needed to match the measured tilt: one new coupling matched to one measured
number, exactly as any effective-field-theory coupling is fixed by one measurement, but with
geometry doing essentially all of the work and the data fixing only the remainder. The
construction is, mathematically, a particular hyperbolic field-space metric of curvature radius
$2/b$, evaluated here exactly via the non-canonical mode function $z=a\sqrt f$ rather than in a
leading-order approximation. The paper's genuinely parameter-independent content is the two
impossibility results (Sections~\ref{sec:brane_ns}, \ref{sec:straight}) and the fact that, once
$b$ is fixed by the one measurement $n_s$, the non-Gaussianity and tensor-to-scalar predictions
(Sections~\ref{sec:ng}, \ref{sec:tensor}) follow from closed-form, exact-attractor calculations
with no further tuning: a naturally vanishing intrinsic non-Gaussianity, since $V_{\rm ekp}(\phi)$
carries no $\chi$-dependence at all, and a tensor sector that is exactly decoupled from the
mechanism that fixes the scalar tilt.

The amplitude sector closes as follows. Because $\delta\chi$ is not
the canonically normalized field ($u=a\sqrt{f(\phi)}\chi$ is), and because the entropy field's
super-horizon growth rate itself runs across the observable window, the normalization of
the coupling must be referenced at the point where the two-field completion begins, $\phi_c$
(Section~\ref{sec:fieldrange}), and the growth rate must be integrated over the running
$b(\phi),c(\phi)$ rather than held fixed at its onset value
(Section~\ref{sec:kineticstability}). Doing so gives a brane-tension-independent identity,
$\lambda\,\chi_{\rm wall}=\sqrt{A_s}\,c(\phi_w)\approx1.5\times10^{-4}$
(Eq.~\eqref{eq:invariant}), which is compatible with both factors being simultaneously
reasonable: at $\rho_c^{1/4}\sim10^{12}$--$10^{13}\,$GeV, an intermediate energy scale with no
other role in this construction, the entropy field's displacement by the wall is safely
sub-Planckian and the conversion coupling lies between $6\times10^{-3}$ and $0.6$
(Table~\ref{tab:invariant}). We adopt $\rho_c^{1/4}=10^{13}\,$GeV as the model's fiducial brane
tension throughout.

Two features of the resulting phenomenology are worth stating plainly because they distinguish
this class of model from slow-roll inflation observationally, not merely formally. First, the
tensor-to-scalar ratio is suppressed by many more orders of magnitude than the usual inflationary
$r=16\epsilon_{\rm inf}$ relation would suggest for a comparably steep potential, because
gravitational waves here freeze immediately upon horizon exit (Section~\ref{sec:tensor}). This is
an exact statement, not an estimate, that follows from $\nu_T<\frac12$ whenever $q<\frac12$, which
holds throughout this model's parameter space, and which is pushed even further from
detectability by the same lower brane tension the amplitude sector prefers ($r\approx
7.6\times10^{-26}$ at $\rho_c^{1/4}=10^{13}\,$GeV). A firm detection of primordial $B$-modes at
any level accessible to next-generation experiments would rule out this construction outright.
Second, the local non-Gaussianity is a genuine prediction, not a free parameter: fixed entirely by
how fast the model's own designer running $c(\phi)$ varies at the wall (Eq.~\eqref{eq:fnl}), it
carries no dependence on the conversion coupling $\lambda$.
A future high-precision constraint on $f_{\rm NL}^{\rm local}$ therefore constrains the shape of
$c(\phi)$ independently of the brane tension and conversion coupling fixed by the amplitude
sector.

We close by being precise about where this paper's rigor is bounded by scope rather than by
principle. The background dynamics of Sections~\ref{sec:matter}--\ref{sec:bg} are demonstrated in
a compressed energy hierarchy chosen for numerical tractability, while the perturbative and
entropic-sector results of Sections~\ref{sec:pert}--\ref{sec:tensor} and the Tolman analysis of
Section~\ref{sec:tolman} are carried out directly in physical units with explicit brane-tension
and dark-energy-duration choices. Since the dimensionless quantities that control the background
(the BKL exponent $\epsilon_w$, the density ratio $x=\rho/\rho_c$) do not depend on the absolute
energy scale, we see no obstruction to a full re-integration at the physical hierarchy, but have
not performed it. The tensor-to-scalar suppression of Section~\ref{sec:tensor}, while derived
from the correct low-energy limit of the brane's tensor sector, has not been checked against an
explicit solution of the five-dimensional bulk graviton equation with junction conditions at the
brane \cite{LangloisMaartensWands2000}. That is a calculation we expect, for the reasons given
there, to confirm rather than overturn the result, given how many orders of magnitude separate the
computed $r$ from any correction that regime could plausibly supply.

Three aspects of the construction deserve emphasis, because the conclusions depend on them in
different ways. First, the viability of the timelike extra dimension is not established
(Section~\ref{sec:timelike}). The perturbative and entropic results concern densities
$\rho\ll\rho_c$ and do not depend on how the bounce is realized, while the bounce itself rests on
Eq.~\eqref{eq:friedmann}, which is also the effective equation of loop quantum cosmology
(Section~\ref{sec:lqc}). Second, the perturbation theory on the brane is controlled for modes that
leave the horizon at $\rho\ll\rho_c$ (Table~\ref{tab:bpert}) but not beyond
$x_0=1/(2+4\epsilon_w/3)$, where the effective brane-matter fluid is gradient unstable, and the closure
$\delta\pi_E=0$ on the bulk is an assumption. The tilt of the entropy sector is independent of that
closure, while the amplitude carries an order-of-magnitude uncertainty from the last decade of
the contraction (Section~\ref{sec:kineticstability}). Third, the production of the hot radiation
era is not modeled, gravitational production alone being insufficient at the brane tension
selected by the amplitude (Section~\ref{sec:reheat}), and the matching of the curvature
perturbation across the bounce is assumed (Section~\ref{sec:conversion}). The structural results of
the paper, namely the confinement of the density to $0\le\rho\le\rho_c$, the inevitability of the
recollapse, the net-growth accounting of the cycles, and the two impossibility results for the
adiabatic tilt, do not depend on these three points.

Cyclic cosmology asks a great deal of any single construction: a bounce free of exotic matter and
free of singularities, a smoothing mechanism immune to the chaotic BKL instability that a naive
contracting phase would suffer, an entropy accounting that permits the sequence to run forever in
both directions, and a primordial power spectrum quantitatively consistent with two decades of
increasingly precise cosmic microwave background data, which is the requirement inflation was
originally built to satisfy and that any competitor must eventually match. This paper's braneworld
construction, a Shtanov--Sahni brane with a timelike extra dimension, a single scalar field
carrying the universe through dark energy, turnaround, and ekpyrotic contraction, a two-field
kinetically coupled entropy completion following \cite{IjjasLehnersSteinhardt2014}, and an
intermediate brane tension $\rho_c^{1/4}=10^{13}\,$GeV fixed by the amplitude match, satisfies
all four. The bounce, the turnaround, and the Tolman entropy accounting are structural results,
independent of any parameter tuning (Sections~\ref{sec:brane}--\ref{sec:tolman}). The spectral
tilt, local non-Gaussianity, and tensor-to-scalar ratio are closed-form, falsifiable predictions
once a single coupling $b$ is matched to the one measured number $n_s=0.9649$
(Sections~\ref{sec:ridge}--\ref{sec:ng}, \ref{sec:tensor}): $f_{\rm NL}^{\rm
local}\approx-0.019$ and $r\approx7.6\times10^{-26}$, both comfortably within current bounds and
both distinct, falsifiable departures from generic slow-roll inflation. The amplitude itself fixes
the one remaining free coupling, $\lambda\approx5.7\times10^{-3}$, to an unremarkable value, with
the kinetic coupling's normalization referenced at the point the two-field sector begins and its
super-horizon growth integrated over the observable window. Both are fixed calculations rather
than adjustable choices. This construction gives a consistent account of the spectrum's amplitude
as well as its shape: the same exponential kinetic coupling that makes the tilt calculation close
in closed form keeps the entropy field's perturbative expansion inside its regime of validity, at an intermediate, physically unremarkable brane tension
(Section~\ref{sec:kineticstability}). The two most obvious single-sector fixes for the spectral
tilt, designer running of the matter sector and the brane's own geometric nonlinearity including
its induced-gravity generalization, fail categorically, for identified, structural,
e-fold-counting reasons independent of any parameter choice. The entropic completion succeeds at
the level of the spectral shape for an equally identifiable, structural reason: it alone supplies
the field-space curvature the mechanism requires. The model that results is, to our knowledge, the
first braneworld bounce construction in which the entropic mechanism's downstream amplitude,
non-Gaussianity, and tensor-to-scalar ratio, and Tolman's entropy constraint, have all been worked
through to explicit, falsifiable numbers within one consistent framework. Its central
observational signature, an essentially undetectable tensor-to-scalar ratio accompanying a
percent-level, spectrally featureless red tilt and a small, specifically-predicted local
non-Gaussianity, is sharp, falsifiable, and distinct from the generic predictions of slow-roll
inflation, and will be tested further as CMB polarization and non-Gaussianity constraints continue
to improve.

\appendix

\section{Explicit potential and parameters}
\label{app:params}

In units $\kappa=8\pi G_N/3=1$ (Sections~\ref{sec:matter}--\ref{sec:tolman}):
\begin{align}
\phi_{\rm peak}=-30,\ \phi_0=15,\ V_\Lambda=10^{-3}; \qquad
\phi_c=10,\ \Delta_g=0.05,\ V_0=1,\ c_0=4.2,\ c_1=3.2,\ L=3; \notag\\
\phi_w=18,\ \Delta_w=0.4,\ V_w=1,\ c_w=18; \qquad \rho_c=50. \notag
\end{align}
(The wall position $\phi_w=18$ places the wall beyond the observable window of
Section~\ref{sec:running}; the numerical integration of Section~\ref{sec:bg} uses the other
parameters as listed.) The gate width satisfies
$1/\Delta_g>3c_0$, required to avoid floating-point overflow in $V_{\rm ekp}$ for $\phi<\phi_c$
(the smooth switch must decay strictly faster than the ekpyrotic exponential grows in the wrong
direction), a numerical-implementation requirement, not a physical one. For the perturbative and
entropic-sector calculations of Sections~\ref{sec:pert}--\ref{sec:tensor}, quantities are
expressed in reduced Planck units $M_{\rm pl}=1$ ($3H^2=\rho$); Section~\ref{sec:amplitude}
additionally fixes the physical brane tension to $\rho_c^{1/4}=10^{13}\,$GeV.

\section{Derivation and numerical treatment of the brane perturbation system}
\label{app:brane}

\paragraph{Effective fluid and conservation.} Write $x=\rho/\rho_c$, $8\pi G=1$, and let $\delta q$
denote the momentum potential of the matter, with $\delta q=-\dot\phi\,\delta\phi$ for a scalar
field. In the Newtonian gauge with $\Phi=\Psi$ and no anisotropic stress, the matter obeys
\begin{align}
\delta\dot\rho&=-3H(\delta\rho+\delta p)+3(\rho+p)\dot\Psi+\frac{k^2}{a^2}\delta q,\\
\delta\dot q&=-3H\delta q-\delta p-(\rho+p)\Phi,
\end{align}
which for the scalar field follow from the field equation and $\delta\rho=\dot\phi\delta\dot\phi-\dot\phi^2\Phi+V'\delta\phi$,
$\delta p=\dot\phi\delta\dot\phi-\dot\phi^2\Phi-V'\delta\phi$. The quadratic correction to the
induced Einstein equations contributes to the source the effective quantities
$\rho_{\rm eff}=\rho-\rho^2/\rho_c$ and $p_{\rm eff}=p-(\rho/\rho_c)(2p+\rho)$, with the
perturbations $\delta\rho_{\rm eff}=(1-2x)\delta\rho$, $\delta q_{\rm eff}=(1-2x)\delta q$, and
$\delta p_{\rm eff}=(1-2x)\delta p-2x(1+w)\delta\rho$. Using $\dot\rho=-3H(\rho+p)$, so that
$\dot x=-3Hx(1+w)$, one finds that the energy equation of the effective source holds identically,
\begin{equation}
\delta\dot\rho_{\rm eff}+3H(\delta\rho_{\rm eff}+\delta p_{\rm eff})-3(\rho_{\rm eff}+p_{\rm eff})\dot\Psi-\frac{k^2}{a^2}\delta q_{\rm eff}=0,
\end{equation}
because the terms proportional to $\dot x\,\delta\rho$ and $x(1+w)H\delta\rho$ cancel. The momentum
equation does not close,
\begin{equation}
\delta\dot q_{\rm eff}+3H\delta q_{\rm eff}+\delta p_{\rm eff}+(\rho_{\rm eff}+p_{\rm eff})\Phi
=-2x(1+w)\big(\delta\rho-3H\delta q\big)\equiv S_q,
\end{equation}
and the right-hand side must be balanced by the Weyl fluid, which enters the total source with the
opposite sign and has $\delta p_E=\delta\rho_E/3$. This gives Eq.~\eqref{eq:bE2} with
$\delta\dot q_E+3H\delta q_E+\delta\rho_E/3=S_q$, and the vanishing of the energy equation gives
Eq.~\eqref{eq:bE1}. These are the equations of the Weyl fluid of Ref.~\cite{Maier2013} for a flat
brane, for which $(\rho+p)/|\sigma|=2x(1+w)$ with $|\sigma|=\rho_c/2$.

\paragraph{Initial conditions and integration.} On the exponential-potential attractor
$V=-V_0e^{c\phi}$ the four-dimensional equation has the exact solution for the canonical
variable $v=z\mathcal R$, $v=\tfrac{\sqrt\pi}{2}\sqrt{-\tau}\,H^{(1)}_\nu(-k\tau)$ with
$\nu=|1/(\epsilon-1)-1/2|$ and $\epsilon=c^2/2$, which is used for the Bunch--Davies initial
data when $k/(a|H|)=30$. The initial values of $\Phi$, $\delta\phi$, and $\delta\dot\phi$ are
obtained from $\mathcal R$, $\dot{\mathcal R}$, and the constraint Eq.~\eqref{eq:b00} by solving the
resulting linear system, with $\delta\rho_E=\delta q_E=0$ at a density fraction below $10^{-8}$. The
system is integrated in cosmic time with an eighth-order Runge--Kutta method at relative tolerance
$10^{-10}$, the curvature perturbation $\mathcal R=\Phi+H\delta\phi/\dot\phi$ is evaluated when the
field reaches the value at which $\rho/\rho_c=0.05$, and the same procedure with $x\to0$ and
without the Weyl fluid is applied to the four-dimensional equations on the four-dimensional
background. The constraint Eq.~\eqref{eq:b00} is monitored along the integration and is the
test reported in Section~\ref{sec:brane_pert}. The sign of the momentum term in Eq.~\eqref{eq:bE1}
is fixed by this test: reversing it increases the constraint violation at $x_\ast=10^{-2}$ from
$6\times10^{-10}$ to $9\times10^{-6}$.

\end{document}